\documentclass[aps, 10pt, prd, notitlepage, twocolumn, superscriptaddress,
               tightenlines, nofootinbib]{revtex4-1}

\usepackage[linktocpage,breaklinks]{hyperref}
\usepackage[usenames,dvipsnames]{xcolor}

\usepackage{aas_macros}
\usepackage{amsmath}
\usepackage{amssymb}
\usepackage{amsfonts}
\usepackage{txfonts}
\usepackage{bm}
\usepackage{stmaryrd}
\usepackage{tensor}
\usepackage[utf8]{inputenc}

\usepackage{graphicx}
\usepackage{epsfig}
\usepackage{epstopdf}

\usepackage{natbib}
\usepackage{hyperref}
\usepackage{cleveref}

\definecolor{mred}{RGB}{127,0,25}
\definecolor{mdgr}{RGB}{51,51,51}
\definecolor{mag}{RGB}{211, 54, 130}

\hypersetup{colorlinks=true,
            citecolor=mred,
            linkcolor=mred,
            urlcolor=mred}

\newcommand{\dd}{{\rm d}}

\newcommand{\vp}{\varphi}

\newcommand{\del}{\partial}

\begin{document}
\title{\sc Neutron star pulse profiles in scalar-tensor theories of gravity}

\begin{abstract}
The observation of the x-ray pulse profile emitted by hot spots on the surface of
neutron stars offers a unique tool to measure the bulk properties of these
objects, including their masses and radii.
The x-ray emission takes place at the star's surface, where the gravitational field is
strong, making these observations an incise probe to examine the curvature of spacetime
generated by these stars.
Motivated by this and the upcoming data releases by x-ray
missions, such as NICER (Neutron star Interior Composition Explorer), we
present a complete toolkit to model pulse profiles of rotating neutron stars
in scalar-tensor gravity.
We find that in this class of theories the presence of the scalar field
affects the pulse profile's overall shape, producing strong deviations from the general relativity expectation.
This finding opens the possibility of potentially using x-ray pulse profile data to
obtain new constraints on scalar-tensor gravity, if the pulse profile is found to be in
agreement with general relativity.

\end{abstract}

\author{\sc{Hector O. Silva}}
\email{hector.okadadasilva@montana.edu}
\affiliation{eXtreme Gravity Institute,
Department of Physics, Montana State University, Bozeman, Montana 59717 USA}

\author{\sc{Nicol\'as Yunes}}
\email{nicolas.yunes@montana.edu}
\affiliation{eXtreme Gravity Institute,
Department of Physics, Montana State University, Bozeman, Montana 59717 USA}

\date{{\today}}

\maketitle

\section{Introduction}
\label{sec:intro}

Neutron stars are some of the most extreme objects in the Universe,
and thus, they serve as a unique laboratory to probe fundamental physics.
Their large masses ($m \approx 1.4\, M_{\odot}$) combined with their small radii
($R \approx 12$ km) result in supranuclear densities at their cores,
whose description challenges our current understanding of matter. The
latter is encoded in the star's equation of state, whose determination is an
outstanding problem in nuclear astrophysics~\cite{Lattimer:2015nhk}.
Moreover, the strong gravitational fields produced by neutron stars result in gravitational potentials
$\sim G m /(R c^2)$ that are nine orders of magnitude
larger than what we can probe on Earth's surface~\cite{Will:2014kxa,Baker:2014zba}.
Therefore, to correctly describe these stars, we must rely on a relativistic
theory of gravity~\cite{Bonolis:2017fdf}.

The leading theory is of course Einstein's theory of general relativity. During its
centennial existence, the theory has shown a remarkable predictive power, being
consistent within all experimental tests carried out so far, ranging from local,
Solar System experiments~\cite{Will:2014kxa}, to the spectacular
detection of gravitational waves by the LIGO/Virgo
collaboration~\cite{Abbott:2016blz,TheLIGOScientific:2016src}.
This consistency with observation is even
more striking when one notes that the theory (unlike most of its alternatives)
does not possess any free parameters that can be tuned to make its predictions
agree with Nature.

Given the success of general relativity, why should we even consider
modifications to it and examine their observational consequences?
The reasons are many, but they can be organized in
two main classes~\cite{Clifton:2011jh,Berti:2015itd,Alexander:2009tp}.
On the observational front, the late time expansion of the
Universe~\cite{Riess:1998cb,Perlmutter:1998np}, the rotation
curve of galaxies~\cite{Sofue:2000jx,Bertone:2016nfn}, the baryon-antibaryon
asymmetry~\cite{Spergel:2003cb,Canetti:2012zc},
and other cosmological observations seem to point at either exotic dark fluids
or modifications to general relativity.
On the theoretical front, the incompatibility of general relativity with quantum
mechanics has prompted many attempts at extensions, from string theory to
loop quantum gravity and other variations.

Can neutron star observations\footnote{We here focus
on properties of {\it individual} stars. Neutron stars have already proven
invaluable tools to constrain deviations of general relativity when either in
binaries~\cite{Damour:2007uf,Wex:2014nva,Kramer:2016kwa} (or
triple~\cite{Archibald:2018oxs}) systems or
more recently binary neutron stars mergers~\cite{TheLIGOScientific:2017qsa}
(see e.g.~\cite{Sakstein:2017xjx,Baker:2017hug,Ezquiaga:2017ekz,Creminelli:2017sry,Bettoni:2016mij}).}
be used to learn about gravity in extreme environments?
A general prediction of modified theories of
gravity is that the bulk properties of the star (e.g. its radius and mass)
are different from those predicted by general relativity.
Tests of gravity in this direction are however
limited by a strong degeneracy problem between the equation of state and the gravity theory:
the modifications of the bulk properties of these stars caused by changes
in the gravitational theory are (often) degenerate with modifications due to
different equations of state~\cite{Glampedakis:2015sua}.

One option to bypass this issue is to focus on electromagnetic phenomena
in the vicinity of neutron stars~\cite{Psaltis:2008bb}. These phenomena include
e.g.~atomic spectral lines~\cite{DeDeo:2003ju},
burst~\cite{Psaltis:2007rv,Glampedakis:2016pes} and quasiperiodic
oscillations~\cite{DeDeo:2004kk,Doneva:2014uma,Pappas:2015npa,Glampedakis:2016pes}.
One can argue that in these scenarios one can in principle probe the {\it exterior spacetime} of the star,
offering a glimpse on possible deviations from general relativity, without
worrying about the intricacies of the stellar interior.

The observation of x-ray waveforms or pulse profiles
from rotating neutron stars is another potentially interesting
phenomenon to consider~\cite{Arzoumanian:2009qn}.
In this scenario, a region of of the neutron star's surface becomes hot
(relative to the rest of the star's surface) generating an x-ray flux modulated
by the star's rotation.
This {\it hot spot} can be formed in a number of situations
(see~\cite{Poutanen:2008pg,Ozel:2012wu,Watts:2016uzu} for reviews).
In accretion-powered pulsars, material is channeled through the magnetic field
lines and heats the star up when it reaches its magnetic poles.
In burst oscillations, a thermonuclear explosion caused by infalling
matter results in a hot spot on the surface of the accreting neutron
star.
In all these cases, the modeling of the resulting waveform, combined with
x-ray observations, allows for the extraction of a number of properties of the
source, including the neutron star's mass and
radius, see e.g.~\cite{Miller:1997if,Weinberg:2000ip,Muno:2002es,Bogdanov:2012md,Poutanen:2003yd,Lo:2013ava,Miller:2014mca,Stevens:2016xiw,Lo:2018hes,Salmi:2018gsn}.

The ongoing NICER (Neutron star Interior Composition Explorer) mission~\cite{2012SPIE.8443E..13G,2014SPIE.9144E..20A,2017NatAs...1..895G} offers a
substantial improvement over the preceding x-ray observatory,
the Rossi X-ray Timing Explorer (RXTE), opening the path to measurements of
stellar masses and radii with unprecedented accuracy -- with immediate implications
to our understanding on the neutron star equation of star.
Given, the scientific potential of NICER, it is natural to ask:
{\it can we use its observations to probe the strong-field regime of gravity}?
In this paper, we take the first necessary steps to find an answer to this question.
We find hints that NICER can indeed be used to probe the strong-field
regime of gravity, but a definite answer will require a plethora of theoretical and
data analysis work that this paper now enables
(see~\cite{Sotani:2017bho,Sotani:2017rrt} for independent recent work in this direction).

\subsection{Executive summary}
We present a complete toolkit to model the x-ray flux from radiating
neutron stars in scalar-tensor gravity, one of the most well-studied and
well-motivated extensions of general relativity~\cite{Chiba:1997ms,Sotiriou:2015lxa}.
This class of theories extends general relativity by introducing a scalar field that
couples to the metric nonminimally, thus violating the strong-equivalence principle.
Our formalism and the resulting toolkit are completely model-independent
within this class of scalar-tensor theories. Moreover, the resulting waveforms
include Doppler shifts, relativistic aberration and
time-delay effects, thus extending~\cite{Sotani:2017rrt}. All of these effects
are critical ingredients that are necessary to develop an accurate pulse-profile model,
and thus, our toolkit now enables a complete data analysis study that will be carried
out in the future.

\begin{figure}[t]
\includegraphics[width=0.48\textwidth]{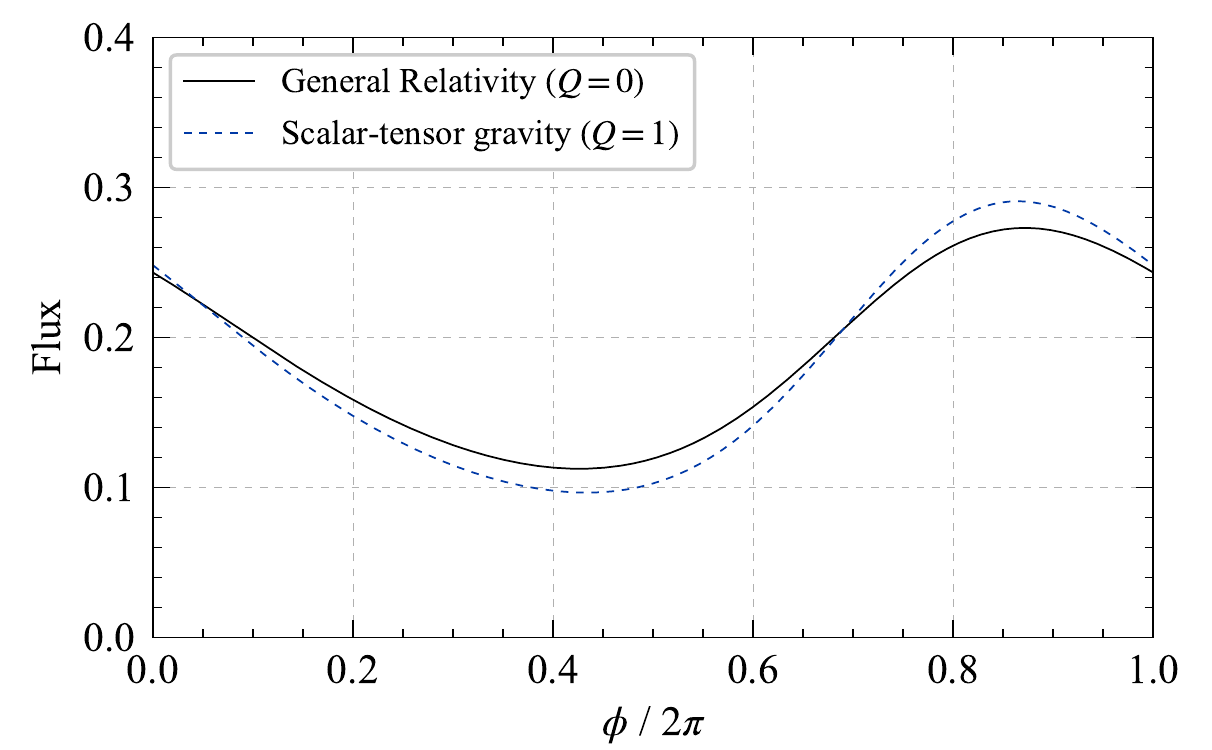}
\caption{An illustrative bolometric flux for a single hot spot over one
revolution of the rotating neutron star in general relativity and scalar-tensor
gravity. The star has $u \equiv 2 G_{\ast} m / (R c^2) = 0.5$ and mass
$m = 1.4\, M_{\odot}$ in both cases. The only difference between the two models
is the presence
of a nonzero scalar charge for the star in scalar-tensor gravity. The magnitude
of the scalar charge is quantified by the scalar charge-to-mass ratio $Q$ (described
in Sec.~\ref{sec:review}) which is zero in general relativity and controls by how
much the spacetime is different from the usual Schwarzschild spacetime.
This particular examples a sample star labeled $c_3$ as discussed in
Sec.~\ref{sec:review} (cf. Table~\ref{tab:stellar_models}).}
\label{fig:pp_example}
\end{figure}

Figure~\ref{fig:pp_example} shows a sample waveform that takes all the
previously mentioned effects into account, illustrating the difference
between the predictions of both theories. As will discuss in
Sec~\ref{sec:pulseprofile}, we find that the presence of the scalar field influences the
exterior spacetime of the neutron star, altering the bending of light and the time-delay
experienced by photons emitted by the star.
The net result of these contributions is a waveform that can be considerably
different in comparison to general relativity's predictions. Such a model now enables,
for the first time, a serious data analysis investigation of whether such waveform differences
can be detected/constrained with data or whether they are degenerate with other
waveform parameters.

The remainder of this paper is organized as follows.
In Sec.~\ref{sec:review}, we briefly review the basic of scalar-tensor gravity,
discussing the properties of the exterior spacetime of neutron stars in this theory.
Next, in Sec.~\ref{sec:pulseprofile}, we present in detail how we construct
our pulse profile model.
Finally, in Sec.~\ref{sec:results} we show a few examples of the pulse
profile of burst oscillations and discuss the modifications introduced by
scalar-tensor gravity. We discuss in details the roles played
by different effects on the overall shape of the resulting waveform.
We close with Sec.~\ref{sec:conclusion}, summarizing our main findings and
discussing some possible extensions and applications of this work.

\section{Overview of scalar-tensor theory}
\label{sec:review}

\subsection{Action and field equation}

Scalar-tensor theories are one of the most widely studied
and well-motivated extensions to general relativity.
In this class of metric theories of gravity, the gravitational interaction
is mediated by an additional scalar field $\vp$. Assuming this
scalar field to be long-ranged (i.e.~massless), the theory
can be described by an action $S_{\ast}$ in the so-called Einstein
frame given by~\cite{Damour:1992we}:
\begin{align}
S_{\ast} &= \frac{c^4}{4 \pi G_{\ast}} \int \frac{\dd^4 x}{c}\sqrt{-g_{\ast}}
\left[ \frac{1}{4}R_{\ast}
-\frac{1}{2}g^{\mu\nu}_{\ast}\nabla_{\mu}\vp\nabla_{\nu}\vp
\right]
\nonumber \\
&\quad+ S_{\rm m}[\psi_{\rm m}, g_{\mu\nu}]\,,
\label{eq:action}
\end{align}
where $c$ is the speed of light in vacuum, and $G_{\ast}$ is the
bare Newtonian gravitational constant.
The action is written in terms
of the Einstein frame metric $g_{\ast\mu\nu}$, with metric determinant
$g_{\ast} \equiv \textrm{det}\,[g_{\ast\mu\nu}]$ and Ricci curvature
scalar $R_{\ast}$.
Matter fields, collectively represented by $\psi_{\rm m}$, are
minimally coupled to the Jordan frame metric
$g_{\mu\nu} \equiv A^2(\vp)g_{\ast\mu\nu}$, with $A(\vp)$
being a conformal factor. As such, clocks and rods in the real world
measure time intervals and distances of $g_{\mu\nu}$ and not of $g_{\ast\mu \nu}$.
The advantage of working in the Einstein frame is that the field equations
are simpler, making it technically easier to derive predictions
for the theory.
However, in the end, we will express the relevant observables
in the Jordan frame.

The field equations, obtained by varying Eq.~\eqref{eq:action} with
respect to $g^{\mu\nu}_{\ast}$ and $\vp$, are
\begin{align}
R_{\ast\mu\nu} &= 2 \del_{\mu}\vp\del_{\nu}\vp
+ \frac{8\pi G_{\ast}}{c^4}
\left( T_{\ast\mu\nu} - \frac{1}{2}T_{\ast}g_{\ast\mu\nu} \right)\,,
\label{eq:b}
\\
\Box_{\ast} \vp &= -\frac{4\pi G_{\ast}}{c^4} \alpha(\vp) T_{\ast}\,,
\label{eq:a}
\end{align}
where $T_{\ast}^{\mu\nu} \equiv (2 c / \sqrt{-g_{\ast}}) (\delta S_{\rm m} / \delta g_{\ast\mu\nu})$
is the Einstein frame energy-momentum tensor and $T_{\ast} \equiv g_{\ast}^{\mu\nu}T_{\ast\mu\nu}$
is its trace. The energy-momentum tensor acts as a source for the scalar field, through the
coupling $\alpha(\vp) \equiv \dd \log A(\vp) / \dd \vp$. The choice of $A(\vp)$
determines a specific scalar-tensor theory and a number of models have been studied in
the literature.

From the definition of $T_{\ast\mu\nu}$, one can derive the following expressions that connect
the Einstein and Jordan frame energy-momentum tensors (and their traces):
\begin{equation}
T_{\mu\nu} = A^2(\vp) T_{\ast\mu\nu}\,,\,\,
T^{\mu\nu} = A^{-6}(\vp)T^{\ast\mu\nu}\,,\,\,
T = A^{-4}(\vp) T_{\ast}\,.
\end{equation}
Finally, the covariant divergences of $T_{\ast}^{\mu\nu}$
and $T^{\mu\nu}$ read
\begin{equation}
\nabla_{\ast\mu} T_{\ast}^{\mu\nu} = \alpha(\vp) T_{\ast} \nabla^{\nu}_{\ast}\vp\,,
\qquad
\nabla_{\mu} T^{\mu\nu} = 0\,.
\end{equation}

\subsection{Exact exterior neutron star spacetimes}
\label{sec:exact}

The pulse profile of a radiating neutron star depends on its
{\it vacuum exterior} spacetime. In general relativity, the exterior spacetime
of a static, spherically symmetric star is described by the Schwarzschild metric
(by Birkhoff's theorem), whose line element is
\begin{align}
\dd s^2_{\rm Sch} &= -\, f_{\rm Sch} \; \dd (ct)^2
+ f_{\rm Sch}^{-1} \; \dd r^2 + r^2 \dd\Omega\,
\label{eq:schw_metric}
\end{align}
in Schwarzschild coordinates $(t,r,\theta,\phi)$.
In this equation, $\dd\Omega = \dd\theta^2 + \sin^2\theta \; \dd \phi$ is the line
element on a unit-sphere and the Schwarzschild factor is defined via
\begin{align}
f_{\rm Sch} \equiv 1 - 2 G m /(c^{2} r)\,,
\end{align}
with $G$ Newton's gravitational constant, and $m$ the gravitational mass of
the star. Thus, the metric is entirely determined by the mass of
the star.

In scalar-tensor gravity, the exterior spacetime of a static, spherically
symmetric star sourcing a nontrivial scalar field in the Einstein frame
is given by the Just spacetime~\cite{Buchdahl:1959nk,Just1959,Coquereaux:1990qs,Damour:1992we}:
\begin{equation}
\label{eq:just-metric}
\dd s_{\ast}^2 =
- f^{b/a} \dd (ct)^2
+ f^{-b/a} \dd \rho^2
+ \rho^2 f^{1 - b / a} \dd \Omega
\,
\end{equation}
in Just coordinates $(t,\rho,\theta,\phi)$~\cite{Just1959}.
In this equation, $\dd\Omega$ is still the line element on an
unit-sphere, $b \equiv 2 G_{\ast} m / c^2$, but now
\begin{equation}
f \equiv 1 - \bar{a}\,.
\end{equation}
with $\bar{a} \equiv {a}/{\rho}$ and $a$ a real constant
with units of length that is related to both the mass of the star $m$
and the strength of the scalar field. The transformation relating
Just and Schwarzschild coordinates is given by~\cite{Damour:1992we}
\begin{equation}
r = \rho (1 - \bar{a})^{(1 - b/a)/2}\,.
\label{eq:sch_just_radiu_rel}
\end{equation}
but this expression is not analytically invertible. Therefore, the Just line
element in Just coordinates cannot in general be transformed to Schwarzschild
coordinates analytically, although it can be done numerically. The Just
line element in the Jordan frame is related with~\eqref{eq:just-metric} as
$\dd s^2 = A^2(\vp) \dd s^2_{\ast}$.

The scalar field configuration sourced by the star has the form
\begin{equation}
\vp = \vp_\infty + (q / a) \log(1 - \bar{a})\,.
\label{eq:scalar_exact}
\end{equation}
Far from the star ($\bar{a} \ll 1$), $\vp \simeq \vp_\infty - q / \rho$
and therefore, $q$ is the scalar charge of the star, while $\vp_\infty$ is
the cosmological background value of $\vp$, which we assume to be
zero\footnote{Rigorously, the present day value of $\vp_\infty$ has to be
determined from the cosmological evolution of the
theory~\cite{Damour:1992kf,Damour:1993id,Anderson:2016aoi,Anderson:2017phb}.}.
We also assume that the conformal factor is such that $A(\vp_\infty) = 1$ asymptotically.
As a consequence, the Einstein and Jordan-frame masses are the same
$m = m_{\ast}$~\cite{Pani:2014jra,Minamitsuji:2016hkk}. Additionally
the gravitational constant measured in the weak-field regime in a
Cavendish experiment is $G = A^2(\vp_\infty)\, G_{\ast} [1 - \alpha(\vp_\infty)^2]$.

The exterior spacetime of a neutron star in scalar-tensor theories is
therefore determined not just by the star's mass $m$ (or equivalently $b$),
but also by the parameters $q$ and $a$. These constants, however, are
not all independent. Instead, they obey the constraint
\begin{equation}
a^2 - b^2 - (2 q)^2 = 0\,.
\label{eq:constraint}
\end{equation}
We can use this relation to elucidate the meaning of the
ratio $a/b$ present in Eq.~\eqref{eq:just-metric}:
\begin{equation}
a/b = \sqrt{(2q/b)^2 + 1} = \sqrt{1 + Q^2}
\label{eq:ab_ratio}
\end{equation}
where we used $b = 2 G_{\ast} m / c^2$ to define
a dimensionless scalar charge-to-mass ratio
\begin{equation}
Q \equiv q c^2 / (G_{\ast} m).
\label{eq:def_Q}
\end{equation}
The exterior spacetime is therefore fully determined by the pair $(m,Q)$, with $Q$
controlling the departure from the Schwarzschild spacetime, which is recovered
when $Q = 0$ (and thus $a/b = 1$).

\subsection{Constraints on scalar-tensor theories and the parameter space of neutron star models}

In practice, few direct observational constraints on the scalar
charge-to-mass ratio $Q$ exist. In~\cite{Horbatsch:2011nh}, Horbatsch
and Burgess developed a conformal factor $A(\vp)$ and
equation-of-state-independent framework which can be used
to directly constrain $Q$ if a sufficient number of
post-Keplerian parameters are known from a binary system.
When applied to the binaries PSR J0737-3039A/B and
PSR B1534+12 they obtained the upper bounds $|Q| = 0.21$ and $|Q| = 0.44$ (at 68$\%$
confidence level) respectively.
We use the latter value to guide an estimate of the largest magnitude of $Q$
allowed from observations; for concreteness we use $Q = 0.5$.

The parameter $Q$, is also constrained by
theoretical considerations. Clearly, for any
given equation of state, if the compactness is large enough, the
neutron star will collapse and form a black hole
(see~\cite{Palenzuela:2015ima,Mendes:2016fby} for studies in particular
scalar-tensor gravity models). An estimate of
this compactness was found by Buchdahl~\cite{Buchdahl:1959zz,Wald:1984rg}
in general relativity [$G m / (R c^2) < 4/9$].
Extending this work to scalar-tensor theories, Tsuchida et al.~\cite{Tsuchida:1998jw}
found that in the Einstein frame, $a$, $b$ and $q$ have their values bounded under
a minimal set of assumption akin to those of Buchdahl~\cite{Wald:1984rg}:
that the star is a perfect fluid, that the density is positive (and monotonically
decreasing) and that the solution matches the Just metric at the star's surface.

We can write this constraint on $Q$ differently by working directly
with the parameters $a$ and $b$ from Eq.~\eqref{eq:ab_ratio}. With this
reparametrization, the result from~\cite{Tsuchida:1998jw}
delimits a domain $\mathcal{D}$ (in a plane spanned by
${\bar a}_{\rm s} = a / \rho_{\rm s}$ and
${\bar b}_{\rm s} \equiv b / \rho_{\rm s}$, where $\rho_{\rm s}$ is the
star's radius in Just coordinates) given by:
\begin{align}
{\bar b}_{\rm s} \leq {\bar a}_{\rm s} \leq 2\sqrt{{\bar b}_{\rm s}} - {\bar b}_{\rm s}\,,
\,\,\, &\textrm{for} \,\,\, 0 \leq {\bar b}_{\rm s} \leq 4\left(3-2\sqrt{2}\right)
\nonumber \\
{\bar b}_{\rm s} \leq {\bar a}_{\rm s} \leq 2\left(\sqrt{2{\bar b}_{\rm s}} - {\bar b}_{\rm s}\right)\,,
\,\,\, &\textrm{for} \,\,\, 4\left(3-2\sqrt{2}\right) \leq {\bar b}_{\rm s} \leq 8/9
\nonumber \\
\textrm{no stars exist} \,, \,\,\, &\textrm{for} \,\,\, {\bar b}_{\rm s} > 8/9
\label{eq:tsuchida}
\end{align}
in which the theory [for any given $A(\varphi)$] admits stellar solutions.
We can thus use these inequalities as a guide to select neutron star models
in scalar-tensor gravity, parametrized by ${\bar a}_{\rm s}$,
${\bar b}_{\rm s}$ and $A_{\rm s}$,
{\it irrespective} of the equation of state of the stellar interior and
of the conformal factor $A(\vp)$.
This makes our analysis {\it as model independent as possible}.

\begin{figure*}
\includegraphics[clip=true,width=0.95\textwidth]{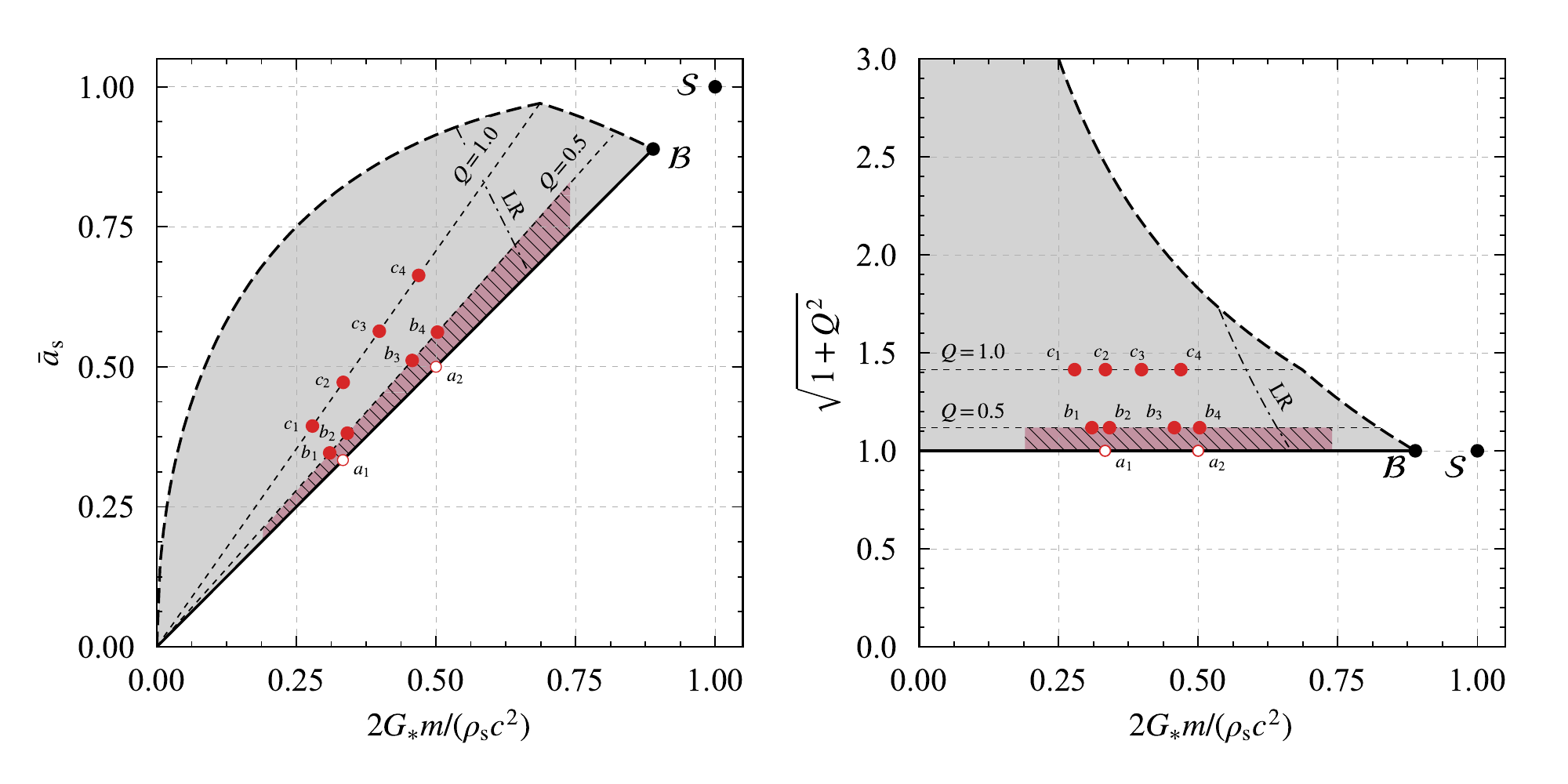}
\caption{The existence domain $\mathcal{D}$ of stars
in scalar-tensor theories parameterized in terms of ${\bar a}_{\rm s}$
and ${\bar b}_{\rm s} [= 2 G_{\ast} m / (\rho_{\rm s} c^2)]$ (left panel)
and in terms of $Q$ and ${\bar b}_{\rm s}$ (right panel).
Stellar solutions exist within the shaded regions of both panels.
General relativistic stars lay on the
solid line ${\bar a}_{\rm s} = {\bar b}_{\rm s}$ and $Q = 0$, with
Newtonian configurations located near the origin of the plane in the left panel
and near the left end in the right panel.
Buchdahl's limit terminates these lines at the point $\mathcal{B} = (8/9, 8/9)$ in the left panel
and $\mathcal{B} = (1,8/9)$ in the right panel, while
black holes are located at the point $\mathcal{S} = (1, 1)$ in both panels.
Moving away from general relativistic line, we enter the realm of stars in scalar-tensor
gravity ($Q \neq 0$).
Their existence is bound from above by the inequalities in
Eq.~\eqref{eq:tsuchida} (dashed curve). The dashed-dotted line (labeled ``LR'')
correspond to the light ring location given by Eqs.~\eqref{eq:lr_bound}.
}
\label{fig:existence}
\end{figure*}
{}
The domain $\cal{D}$ is shown by the shaded region in Fig.~\ref{fig:existence}.
The left panel shows the inequalities in Eq.~\eqref{eq:tsuchida}, while in the
right panel we re-express them in terms of $a/b = {\bar a}_{s} / {\bar b}_{s}
= \sqrt{1 + Q^2}$ instead of ${\bar a}_{s}$.
For reference, we also shows lines of constant $Q$, with the solid
line corresponding to general relativity with $Q=0$.
Black holes are shown by the point labeled $\mathcal{S}$, while ``Newtonian''
solutions (with very small compactness) are at the origin of the plane (left-panel)
or toward the left-end (right-panel). In virtue of
the no-hair theorem of Sotiriou and Faraoni~\cite{Sotiriou:2011dz}, black holes have
zero scalar charge $q = 0$ (thus $Q = 0$, $a = b$) and compactness
$u / 2 = G_{\ast} m / (R c^2) = 0.5$
in theories described by the action in Eq.~\eqref{eq:action}.

The hatched portion of the plane {\it qualitatively} represents the region in
which neutron stars could exist. Assuming small deviations from general relativity
(i.e. small $Q$ as justified previously), we expect
the conformal factor evaluated at the star's surface to be of order unity
$A_{\rm s} \equiv A[\vp(\rho_{\rm s)}] \approx 1$,
hence $R \approx \rho_{\rm s}$, where $R$ is the circumferential, Jordan-frame
radius of the star. Astronomical observations show that neutron stars have
masses approximately in the $m \in [0.9, 2.0]$ $M_{\odot}$ and radii
$R \in [8, 14]$ km ranges, which translate into
$u \approx b_{\rm s} \in [0.19, 0.74]$~\cite{Ozel:2016oaf,Miller:2016pom}.
Therefore, the region between the general relativistic solutions and the constant
$Q = 0.5$ line delimits the region where neutron stars could exist.

An additional region of interest arises from studying when the photon
light ring (or light rings) of the Just metric is inside the radius of the star.
This allows a photon emitted tangentially from the surface of the star
to reach the observer~\cite{Pechenick:1983}. Otherwise, if the light ring is outside of
the star, then tangential photons would come back to hit the surface, and thus, they
would not reach the observer.  In Appendix \ref{app:lr}, we show that
this occurs when
\begin{equation}
{\bar a}_{\rm s} \leq 2(1 - {\bar b}_{\rm s})\,,
\quad \textrm{or} \quad
({\bar b}_{\rm s}/2) (2 + \sqrt{1+Q^2}) \leq 1\,,
\label{eq:lr_bound}
\end{equation}
which is shown by dash-dotted lines in Fig.~\ref{fig:existence}.

Due to the large parameter space available for stellar models in
scalar-tensor gravity we must make a few sensible requirements for choosing
our illustrative stellar models. All stars in our catalog:
(i) obey the constraints~\eqref{eq:tsuchida};
(ii) have a canonical mass of $m = 1.4$ $M_{\odot}$;
(iii) have twice the compactness of $u = \{1/3, 1/2\}$~\footnote{These are
illustrative values used by Poutanen et al.~\cite{Poutanen:2006hw}
which we will use to validate the numerical implementation of our pulse profile
code in  Sec.~\ref{sec:results}}.
In this way, our stellar models in scalar-tensor gravity are
``doppelg\"angers'' of their general relativistic models counterparts:
they have the same gravitational mass $m$ and (Jordan-frame) areal radius
$R$ -- the only difference being the presence of a nonzero scalar charge.
Although we expect $A_{\rm s} \approx 1$, there is no particular
reason to chose it either larger or smaller than unity -- we thus
consider both possibilities.

Under these choices we group them in three classes according to their
value of $Q$.
\begin{itemize}
\item Models $a_{i}$ represent general relativistic stars ($Q=0$).
\item Models $b_i$ represent stars with $Q = 0.5$. To determine
$\rho_{\rm s}$ we assume that the conformal factor evaluated at
the surface $A_{\rm s}$ is $1.0 \pm 0.05$.
\item Models $c_i$ represent stars with $Q = 1.0$. To determine
$\rho_{\rm s}$ we assume that $A_{\rm s}$ is $1.0 \pm 0.1$.
\end{itemize}
The parameters of these stellar models are listed in Table~\ref{tab:stellar_models}.
Given the constraint on $\vert Q \vert$~\cite{Horbatsch:2011nh},
why should we consider values as large as unity? The reasons are twofold. First,
regardless of the constraint, it is of theoretical interest to investigate how
much (and precisely how) this new parameter affects the pulse profile. This question cannot
be answered if we restrict ourselves to $Q \approx 0$. Second, if it were the case
that even for a maximal value of $Q$ in the range of reasonable values (say even for $Q=1$),
the impact of the scalar charge on the pulse profile is minimal, then there would little motivation
to attempt to constraint this theory using pulse profile observations.

\begin{table}
\begin{tabular}{c c c c c c c c c}

\hline
\hline
Name & $u$ & $m/M_{\odot}$ & $R/{\rm km}$ & $\rho_{\rm s}/{\rm km}$
& $A_{\rm s}$ & ${\bar a}_{\rm s}$  &  ${\bar b}_{\rm s}$ & $Q$ \\
\hline
$a_{1}$  &  0.333  &  1.4  &  12.40  &  12.40  &  1.00  &  0.333 &  0.333 &  0.0  \\
$a_{2}$  &  0.500  &  1.4  &  8.269  &  8.269  &  1.00  &  0.500 &  0.500 &  0.0  \\
\hline
$b_{1}$  &  0.333  &  1.4  &  12.40  &  13.35  &  0.95  & 0.346 &  0.309  &  0.5  \\
$b_{2}$  &  0.333  &  1.4  &  12.40  &  12.12  &  1.05  & 0.381 &  0.341  &  0.5  \\
$b_{3}$  &  0.500  &  1.4  &  8.269  &  9.040  &  0.95  & 0.511 &  0.457  &  0.5  \\
$b_{4}$  &  0.500  &  1.4  &  8.269  &  8.227  &  1.05  & 0.562 &  0.503  &  0.5  \\
\hline
$c_{1}$  &  0.333  &  1.4  &  12.40  &  14.83  &  0.90  & 0.394 &  0.279  &  1.0  \\
$c_{2}$  &  0.333  &  1.4  &  12.40  &  12.38  &  1.10  & 0.472 &  0.334  &  1.0  \\
$c_{3}$  &  0.500  &  1.4  &  8.269  &  10.37  &  0.90  & 0.564 &  0.398  &  1.0  \\
$c_{4}$  &  0.500  &  1.4  &  8.269  &  8.817  &  1.10  & 0.663 &  0.469  &  1.0  \\
\hline
\hline
\end{tabular}
\caption{{\it Stellar models.} We list the properties of the stellar
models which we used to compute the pulse profiles. The parameters are:
$u$ (twice the compactness),
$m$ (the gravitational mass), $R$ (Jordan-frame radius in
Schwarzschild coordinates), $\rho_{\rm s}$ (Einstein-frame radius in
Just coordinates), $A_{\rm s}$ (conformal factor evaluated at Einstein-frame
radius in Just coordinate $\rho_{\rm s}$),
$\bar{a}_{\rm s} = a / \rho_{\rm s}$,
$\bar{b}_{\rm s} = b / \rho_{\rm s}$,
$Q$ (scalar charge-to-mass ratio). The conformal factor $A_{\rm s}$
is necessary to translate radii between Einstein to Jordan frames.}
\label{tab:stellar_models}
\end{table}

\subsection{Connection to specific scalar-tensor gravity models}

Although our discussion so far has been fully model independent,
one can easily determine the parameters $(m, Q)$
within a specific scalar-tensor model. We here briefly explain how this
can be done once a specific functional form of $A(\varphi)$ is chosen.

The first step consists of integrating the stellar structure equations of a static,
spherically symmetric neutron star in scalar-tensor theory, which generalize
the Tolman-Oppenheimer-Volkoff equations from general relativity.
These equations are derived from the field equations
[Eqs.~\eqref{eq:b} and~\eqref{eq:a}] assuming matter to be
described by a perfect fluid.
The explicit form of these equations and a description of the
numerical algorithm used to integrate them can be found, e.g.
in~\cite{Damour:1993hw,Harada:1997mr,Silva:2014fca}.

Once a stellar model has been constructed, the
quantities $q$, $m$, $R$ and $A_{\rm s}$ are all known.
From $m$, we immediately determine $b$. Using Eq.~\eqref{eq:constraint}
we find $a$, while from Eq.~\eqref{eq:sch_just_radiu_rel} we obtain
$\rho_{\rm s}$. At last, $Q$ is obtained from Eq.~\eqref{eq:def_Q}.
With these values, the Just spacetime is completely determined.

\section{Pulse profile modeling}
\label{sec:pulseprofile}

Having presented an overview of scalar-tensor gravity, the properties
of the exterior neutron star spacetime and some generic properties of
neutron stars in this theory, we are now ready to develop a pulse profile model.
We start with a description of our assumptions, followed by an analysis
of geodesic motion in the Just spacetime and we close by constructing
the pulse profile model.

\subsection{Assumptions and model parameters}

Several models have been developed to study the pulse profile
of radiating, rotating neutron stars, since the pioneering
work by Pechenick et al.~\cite{Pechenick:1983}
(see~\cite{Poutanen:2008pg,Ozel:2012wu,Watts:2016uzu} for reviews).
Given that this is one of the first detailed study of pulse profiles in
non-GR theories, we will make a series of simplifying physical assumptions,
to be relaxed in future work. In this section, we lay down and
justify these assumptions.

\begin{itemize}
\item [(i)] {\it The stellar model, its exterior spacetime and photon
geodesics.} We assume a spherically symmetric neutron star
described by a solution of the field equations of scalar-tensor
theory with a perfect fluid.
The exterior spacetime (Sec.~\ref{sec:exact}) is described by two
parameters: $Q$ and $m$. The ratio of $a/b$
determines the scalar charge-to-mass ratio $Q$
[cf.~Eqs~\eqref{eq:ab_ratio} and~\eqref{eq:def_Q}].
Photons move on geodesics of the Just spacetime in the Jordan frame~\eqref{eq:just-metric}.

\item [(ii)] {\it Hot spot.} We assume that the star has a small (relative to
the size of the star), uniformly radiating, hot spot on its surface,
located at a polar angle $\theta_{\rm s}$.
Analytical estimates~\cite{Baubock:2015ixa} and numerical
simulations~\cite{Bai:2009wf} in general relativity provide evidence that the
hot spot size has a small impact on the pulse profile observed, provided the spot
is small enough.
In reality, the spot size is likely not sufficiently small for this effect to be
ignorable, but here, for simplicity, we will assume it to be infinitesimal in size and
that the rest of the star is dark, assumptions that can be lifted in the future.
We assume that the scalar field does not affect either the process of
generation of radiation nor its spectral properties.

\item [(iii)] {\it Rotation and special relativistic effects.} The star
rotates with a constant frequency $\nu$ and we neglect the
effects of frame-dragging on the emitted radiation. These effects
have been shown to be small (in general relativity)
and we expect them to be equally small in scalar-tensor
gravity~\cite{Sotani:2012eb}.
At high rotation frequencies ($\nu \gtrsim 300$ Hz), the pulse profile is
mostly affected by the quadrupolar deformation of the
star~\cite{Braje:2000qb,Cadeau:2004gm,Cadeau:2006dc,Morsink:2007tv,Psaltis:2013zja,Nattila:2017hdb}
which, however, we do not include.
However, we do include special relativistic effects of
Doppler boost and aberration due to the rapid motion of the hot spot.
That is, we work in a ``Just-Doppler'' approximation, in
analogy with the ``Schwarzschild plus Doppler'' approximation
introduced by Miller and Lamb~\cite{Miller:1997if}
who worked with the Schwarzschild spacetime
(see also Refs.~\cite{Poutanen:2003yd,Poutanen:2006hw}).

\item [(iv)] {\it Photon time delay.} Photons emitted at different rotation phases
$\phi_{\rm s} \equiv 2 \pi \nu t$ take different paths to reach the
observer, resulting in a time-delay $\Delta t$ (Sec.~\ref{sec:timedelay}). This
time-delay forces the observed phase of the pulse $\phi_{\rm obs}$ to be shifted
with respect to the star's rotational phase as
$\phi_{\rm obs} = \phi_{\rm s} + 2 \pi \nu \Delta t$.
This effect is only appreciable for fast rotators: for a rotational period
of $P = 1.5$ ms, this induces at most a $5$\% correction to the
photon arrival phase in general relativity~\cite{Poutanen:2006hw}.
We do include this effect for completeness.

\item [(v)] {\it Observer.} The observer is located a distance $D$ from the
star, at an inclination $\iota_{\rm o}$ relative to the star's rotation axis.
We express the physical observables in the Jordan frame, which are the
quantities measured by the observer's rods and clocks.
\end{itemize}

\begin{figure}[t]
\includegraphics[width=0.45\textwidth]{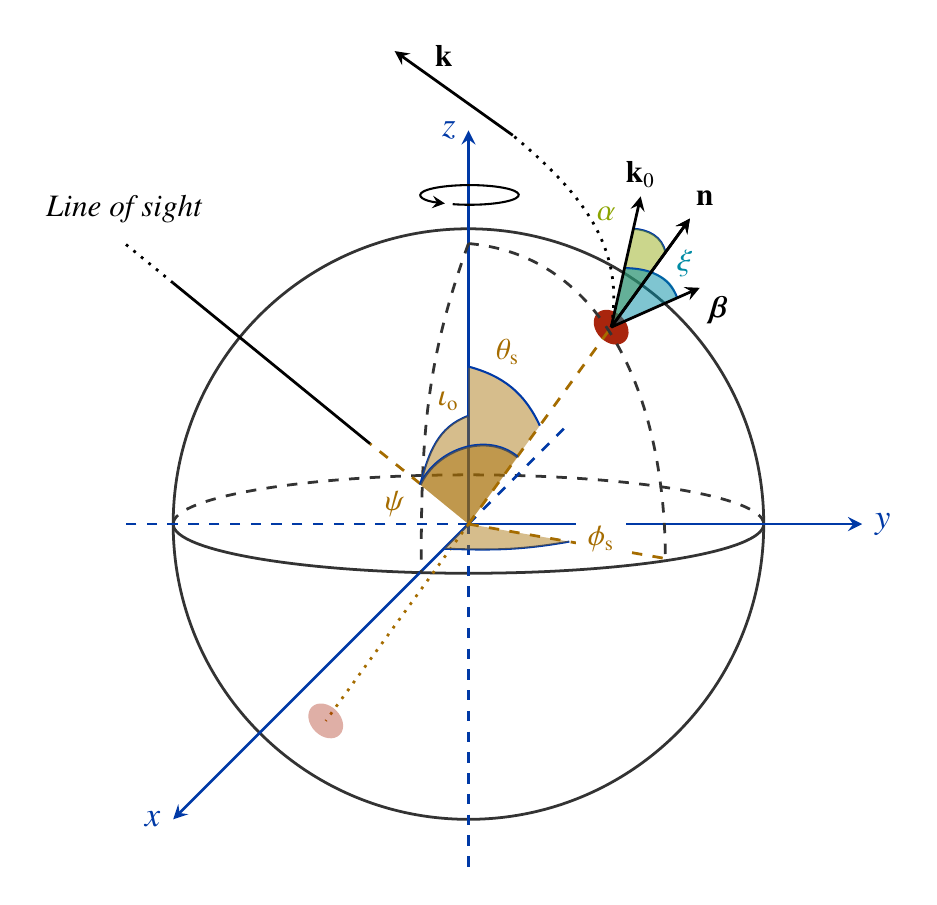}
\caption{Geometry of a rotating neutron star with hot spots.
The star is here depicted as rotating about the positive $z$ axis, with two
antipodal hot spots  in red, and a variety of angles defined in the text.
The line of sight and the vector ${\bm k}$ lay on the same plane which
passes by the origin on the figure.
}
\label{fig:diagram}
\end{figure}

The geometry of the problem is depicted in Fig.~\ref{fig:diagram}.
The hot spot is located at the stellar surface at polar angle $\theta_{\rm s}$
and its instantaneous position is described by a radial unit vector ${\bm n}$.
The unit vector ${\bm n}$ and the line of sight make an angle $\psi$.
The trajectory of emitted photons can be described by a unit vector
${\bm k}_0$, which makes an angle $\alpha$ with respect to the normal ${\bm n}$.
We are interested in the photons whose trajectories (after
experiencing gravitational light bending) are along the line of sight of
the observer. These photon trajectories are depicted by the unit vector
${\bm k}$.
Finally, ${\bm \beta}$ is a vector, perpendicular to ${\bm n}$ and
tangential to the (instantaneous) direction in which the hot spot moves.
We denote by $\xi$ its angle with respect to ${\bm k}_0$.
When present, an antipodal hot spot is located at
$\theta =  \pi - \theta_{\rm s}$ and
$\phi = \pi + \phi_{\rm s}$.

With this at hand, the plan for the remainder of this section is as follows:
(i) to study geodesic motion in the Just spacetime to obtain a relationship
between the angles $\psi$ and $\alpha$, which will tell us which photons
reach the observer, and (ii) to obtain relations for the angles
$\iota_{\rm o}$, $\theta_{\rm s}$, $\phi_{\rm s}$, $\alpha$ and $\xi$ which
we will use to construct the flux measured by the observer, including
special relativistic effects caused by the rapid motion of the hot spot.

\subsection{Null geodesic motion in the Just spacetime}
\label{sec:geodesic}

To construct the bolometric flux measured by an observer, the first
step is to analyze the photon's geodesic motion in the Just metric.
In particular, the goal of this section is to obtain an expression
for the angle $\psi$, shown in Fig.~\ref{fig:diagram}.

We define the Lagrangian
\begin{equation}
L
= g_{\mu\nu} p^{\mu} p^{\nu}\,,
\end{equation}
where the photon's Jordan-frame four-momenta is defined as
$p^{\alpha} \equiv \dd x^{\alpha} / \dd \lambda$
[with $x^{\mu} = (ct,\rho,\theta,\phi)$ the four-trajectory of the photon],
where $\lambda$ is an affine parameter.

Because of spherical symmetry, the photon geodesic is confined
to a constant $\theta$-plane, which we take to be the equatorial plane
($\theta = \pi / 2$).
Using the line element in Eq.~\eqref{eq:just-metric}
to construct $L$, the equations of motion can be obtained
using the Euler-Lagrange equations and the null constraint
$g_{\mu\nu} p^{\mu} p^{\nu} = 0$:
\begin{subequations}
\begin{align}
\dd t / \dd \lambda &=  A^{-2} \varepsilon f^{-b/a}\,,
\label{eq:tdot}
\\
(\dd \rho / \dd \lambda)^2 &=  A^{-4}
\left[c^2\varepsilon^2
- (h/\rho)^2 f^{2b/a - 1}\right]\,,
\label{eq:rhodot}
\\
\dd \theta / \dd \lambda &= 0\,,
\label{eq:thetadot}
\\
\dd \psi / \dd \lambda &= A^{-2} (h / \rho^2) f^{b/a - 1}\,.
\label{eq:psidot}
\end{align}
\end{subequations}
where $\varepsilon$ and $h$ are constants of motion,
associated with energy and angular momentum  respectively.
Combining Eqs.~\eqref{eq:tdot} and~\eqref{eq:rhodot} we obtain:
\begin{equation}
\dd\psi / \dd \rho
= \rho^{-2} f^{b/a - 1}
\left[\sigma^{-2} - \rho^{-2} f^{2b/a - 1} \right]^{-1/2}
\end{equation}
where $\sigma \equiv h / (\varepsilon c)$
is the impact parameter.

\begin{figure}[t]
\includegraphics[width=0.45\textwidth]{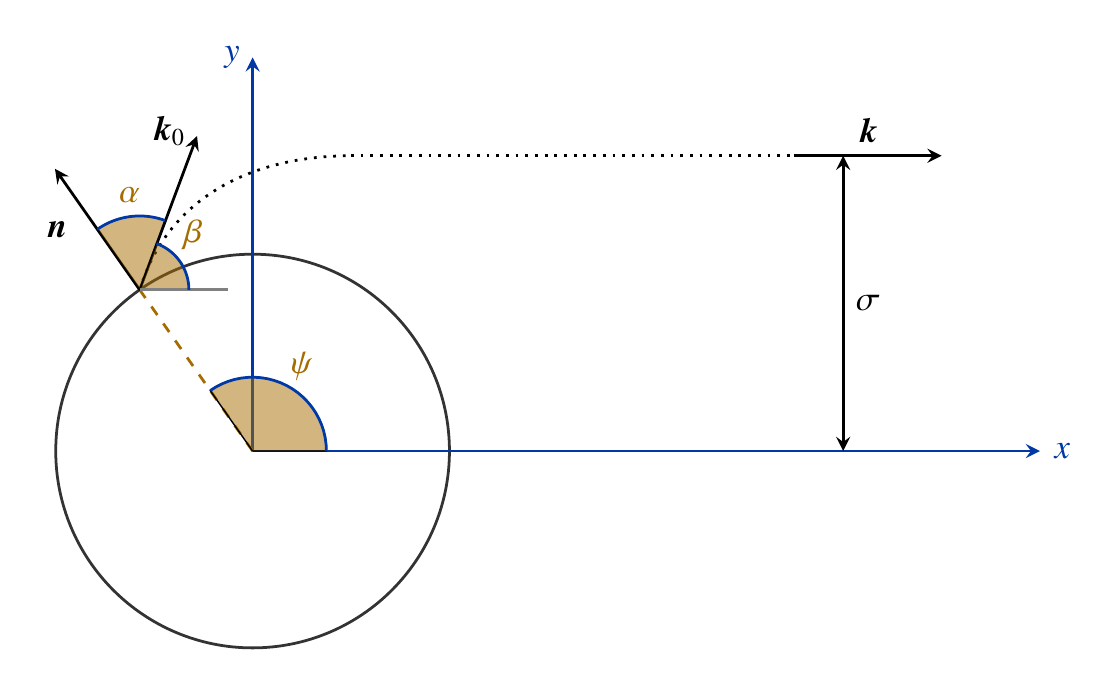}
\caption{Bird's eye view of the rotating neutron star. The photon is
emitted along a unit vector ${\bm k}_0$ with an angle $\alpha$ with
respect to a normal unit vector ${\bm n}$, from a
small hot spot located at $\psi$ with respect to the line of sight.
The star's gravitational field changes the photon trajectory by a
bending angle $\beta \equiv \psi - \alpha$, seen to arrive
with an impact parameter $\sigma$ along the unit vector ${\bm k}$.}
\label{fig:diagram_planar}
\end{figure}

Let us now solve for the angle $\psi$ in integral form.
The impact parameter can be eliminated in
favor of $\alpha$ by noticing that the angle between $p^{\psi}$ and $p^{\,\rho}$
at $\rho = \rho_{\rm s}$ is $\tan \alpha = [p^{\psi} p_{\psi} /
(p^{\,\rho} p_{\rho})]^{1/2}$~\cite{Beloborodov:2002mr} (see Fig.~\ref{fig:diagram_planar}).
Using Eqs.~\eqref{eq:rhodot} and~\eqref{eq:psidot}, we find
\begin{equation}
\sin \alpha = (\sigma / \rho_{\rm s}) (1 - {\bar a}_{\rm s})^{b/a - 1/2}\,,
\label{eq:angle_alpha}
\end{equation}
which in turn results in
\begin{align}
\psi &= \sin\alpha \int_0^1 \dd y \, (1 - {\bar a}_{\rm s} y)^{b/a - 1}
\nonumber \\
&\quad \times
\left[(1 - {\bar a}_{\rm s})^{2b/a - 1}
- y^2 \sin^2\alpha \; (1 - {\bar a}_{\rm s} y)^{2b/a - 1}
\right]^{-1/2}\,.
\end{align}
where we have defined $y \equiv \rho_{\rm s} / \rho$.

This integral has a singularity whenever $\sin \alpha = 1$
and $y \to 1$. The singularity is power-law integrable and
can be removed introducing the new variable
$x = \sqrt{1 - y}$~\cite{Press:2007:NRE:1403886,Lo:2013ava}.
In terms of $x$, we obtain
\begin{align}
\psi &= 2\sin\alpha \int_0^1 \dd x\,
x\, [1 - {\bar a}_{\rm s}(1-x^2)]^{b/a - 1}
\nonumber \\
&\quad \times
\left\{
(1 - {\bar a}_{\rm s})^{2b/a - 1}
\right.
\nonumber \\
&\quad \left.
-\, (1-x^2)^2 \left[1 - {\bar a}_{\rm s}(1 - x^2) \right]^{2b/a - 1}
\sin^2\alpha \right\}^{-1/2}\,.
\label{eq:psi_final}
\end{align}
In the general relativity limit ($a/b = 1$), this result agrees
with the literature~\cite{Poutanen:2003yd,Poutanen:2006hw,Lo:2013ava}.
In the Newtonian limit ($a_{\rm s} = 0$), $\psi = \alpha$
and the bending angle $\beta \equiv \psi - \alpha$
(cf. Fig.~\ref{fig:diagram_planar}) is zero.

Figure~\ref{fig:cospsi} shows $\psi$ calculated in the range
$\alpha \in [0, \pi/2]$, for the stellar models in Table~\ref{tab:stellar_models}.
In the general relativistic limit ($Q=0$ or $a/b=1$) our results agree with~\cite{Poutanen:2006hw}.
A nonzero scalar charge has the effect of increasing/decreasing $\cos\psi$ relative to general relativity
depending on whether the conformal factor at the stellar surface is smaller/larger than unity.
This enhancement/suppression is a manifestation of spacetime becoming compressed/stretched in the star's vicinity in scalar-tensor gravity because of the conformal factor.
This effect is more salient as $\alpha \to \pi/2$ and is virtually
negligible for small $\alpha$.

\subsection{Visible fraction of the star and visibility conditions}

The light ray that defines the star's visible area
is the one emitted at an angle $\alpha = \pi / 2$ (i.e.
tangent to a vector normal to the surface) from a
position specified by an angle $\psi$, measured
with respect to the line of sight of the observer (cf. Fig.~\ref{fig:diagram_planar}).
In flat spacetime, this means that the visibility condition is
\begin{equation}
\cos\psi > 0. \qquad \textrm{flat spacetime}
\end{equation}
Due to light bending, however, we can see more than half
of a spherically symmetric neutron star, i.e.~a photon emitted with
$\cos\psi < 0$ can reach the observer at spatial infinity.
However, there is a critical value
\begin{equation}
\cos\psi_{\rm c} < 0, \qquad \textrm{curved spacetime}
\label{eq:visibility}
\end{equation}
which determines the last visible ring of the star. The hot spot
is visible if
\begin{equation}
\cos\psi > \cos\psi_{\rm c}. \qquad \textrm{visibility condition}
\end{equation}
This value can be calculated numerically using Eq.~\eqref{eq:psi_final}
with $\alpha = \pi/2$, i.e.
$\psi_{\rm c} \equiv \psi(\alpha = \pi/2)$.

\begin{figure}[t]
\includegraphics[width=0.48\textwidth]{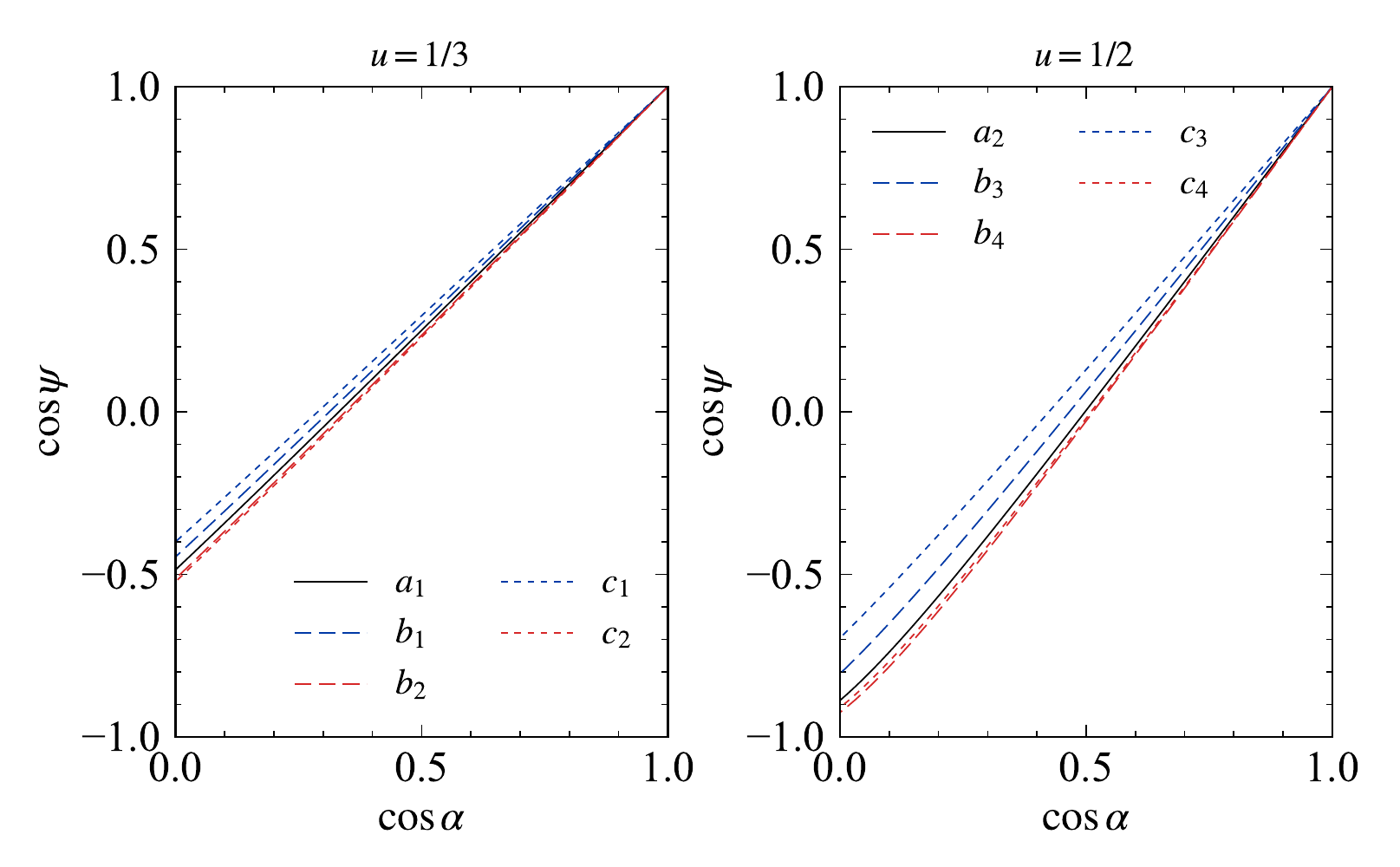}
\qquad
\caption{Effect of the scalar charge on $\psi$ as a function of $\alpha$.
The scalar charge affects $\psi$ mostly at $\alpha \approx \pi/2$, while
for $\alpha \approx 0$ its influence is negligible.
}
\label{fig:cospsi}
\end{figure}

We define the visible fraction of the surface as the ratio:
\begin{align}
\delta f &\equiv
\frac{A_{\rm s}^2\, \rho_{\rm s}^2 (1 - {\bar a}_{\rm s})^{1-b/a}\int_0^{\psi_{\rm c}}\int_0^{2\pi}
\dd\psi'\, \dd\phi \sin\psi' }{4\pi A_{\rm s}^2\,\rho_{\rm s}^2 (1 - {\bar a}_{\rm s})^{1-b/a}}
\nonumber \\
&=
\frac{1-\cos\psi_{\rm c}}{2}.
\end{align}
In the Newtonian limit, $\psi_{\rm c} = \pi/2$ and
$\delta f = 1/2$, i.e. half of the star is visible
as expected.
For relativistic stars, this number increases since
$\cos\psi_{\rm c} < 0$ in general,
as shown in Fig.~\ref{fig:cospsi}. As we have seen in scalar-tensor gravity, this
effect can become larger or smaller in comparison to general relativity depending
on whether the conformal factor at the stellar surface is smaller or larger than unity. For
instance, when $u = 0.5$, $\delta f \approx 0.94$ in general relativity (model $a_{2}$ in Table~\ref{tab:stellar_models}),
while it becomes $\delta f \approx 0.85,\, 0.95$ in scalar-tensor gravity for $A_{s} = 0.9,\, 1.1$
and large $Q=1$ (models $c_3$ and $c_4$ in Table~\ref{tab:stellar_models}).

The above considerations implicitly assume that the light ring
of the star is inside the stellar radius. When this is the case,
tangential photons (ie.~those emitted with $\alpha = \pi/2$) do escape
to spatial infinity. However, if the light ring is outside of the stellar surface,
then tangential photons will not escape to spatial infinity. Instead,
there will be some maximum $\alpha$ for which photons do reach the observer,
and this maximum angle will now need to be used to define the visible fraction
of the star. In this paper, however, for all the stellar models we consider,
the light ring is inside of the stellar surface, and thus, the analysis of
the visible fraction presented above apply.

\subsection{Photon time-delay}
\label{sec:timedelay}

Just as curvature deflects photons around the star, it also slows them down
through the Shapiro time delay effect. The magnitude of this effect can
be calculated using Eqs.~\eqref{eq:tdot} and~\eqref{eq:rhodot}, which gives
\begin{equation}
t = (1/c)\int^{\infty}_{\rho_{\rm s}}
\dd\rho\,
f^{-b/a}\left[1 - (\sigma/\rho)^2 f^{2b/a - 1} \right]^{-1/2}\,,
\end{equation}
and from which we define the time delay as
\begin{equation}
\Delta t \equiv t(\sigma) - t(\sigma=0),
\label{eq:def_deltat}
\end{equation}
defined with respect to a photon emitted directly towards the observer
(i.e. zero impact parameter $\sigma = 0$).
As in the case for $\psi$, we can write the time delay integral
in terms of the emission angle $\sin\alpha$ and introduce the same
integration variables $\rho_{\rm s}/\rho = y = 1 - x^2$, with the
latter to remove (again) a singularity.
The final expression becomes:
\begin{align}
\Delta t &=({2\rho_{\rm s}}/{c})
\int^{1}_0 \dd x \, x \,[1 - {\bar a}_{\rm s}(1- x^2) ]^{-b/a}\,(1-x^2)^{-2}
\nonumber \\
&\quad \times \left\llbracket
\left\{
1 - (1-x^2)^2 (1- {\bar a}_{\rm s} )^{1-2b/a}
\right. \right.
\nonumber \\
&\quad \left. \left. \times \left[ 1 - {\bar a}_{\rm s} (1-x^2)\right]^{2b/a - 1} \sin^2\alpha
\right\}^{-1/2}
- 1 \right\rrbracket\,,
\label{eq:time_delay_final}
\end{align}
which reduces to the general relativistic expression when
$a/b = 1$ or $Q=0$~\cite{Lo:2013ava}.

\begin{figure}[t]
\includegraphics[width=0.48\textwidth]{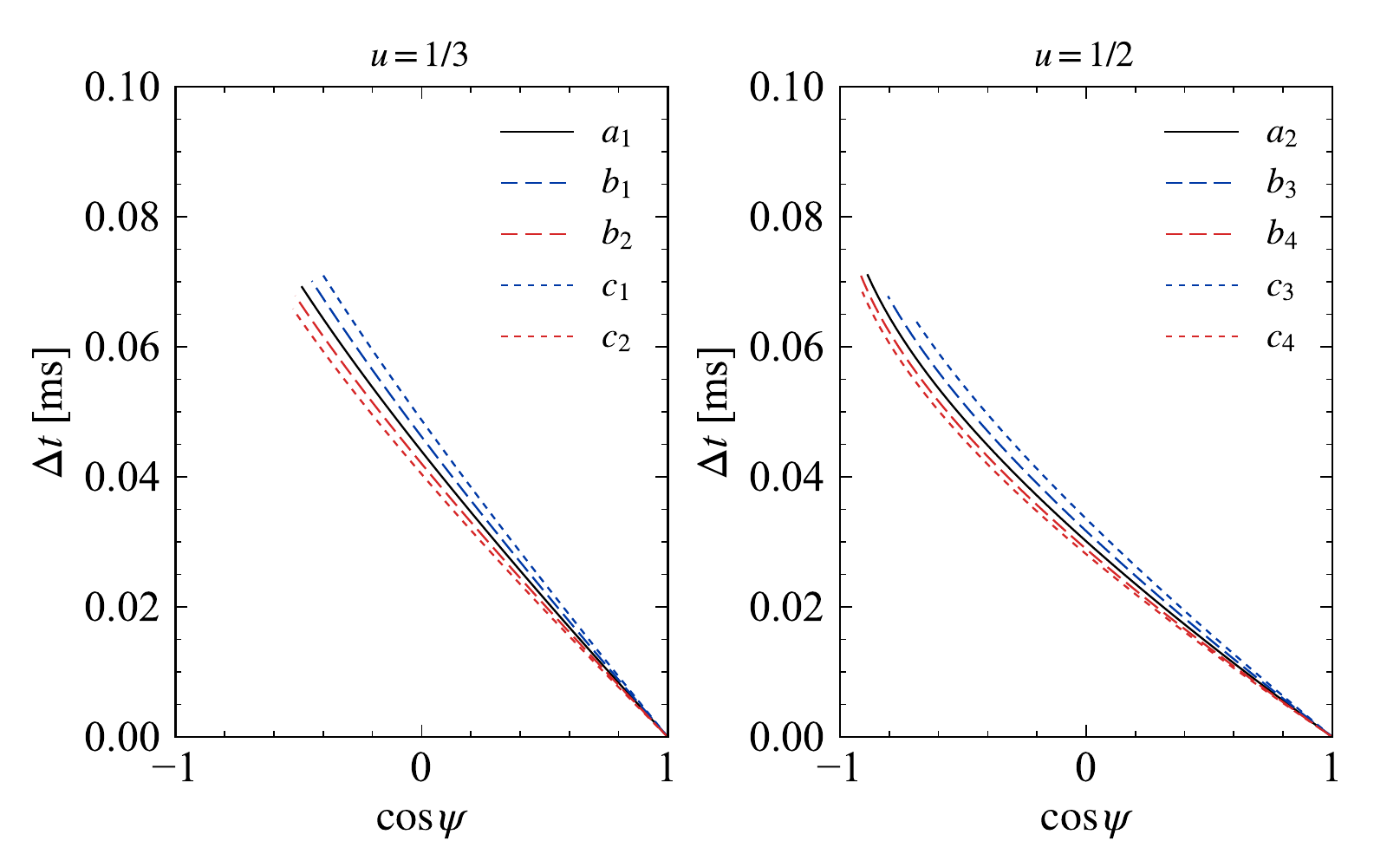}
\caption{Effect of the scalar charge on the time-delay of
photons as they escape the neutron star. As in the case of the angle $\psi$, the effect is
largest when $\alpha$ approaches $\pi / 2$, while it is negligible as
$a \to 0$.
}
\label{fig:timedelay}
\end{figure}

Figure~\ref{fig:timedelay} shows the time-delay $\Delta t$
as a function of angle $\psi$. Our results agree with~\cite{Poutanen:2006hw}
in the limit $a/b = 1$ (or $Q=0$). In scalar-tensor theory,
this delay is weakly dependent on the mass, as also observed in general relativity by~\cite{Poutanen:2006hw},
having a maximum value of $\Delta t \approx 0.07$ ms irrespective of the
$Q$. The deviations from general relativity increase for larger values of $Q$.
Similarly to the calculation of $\psi$, this increase is largest when
$\psi \to \pi / 2$.

\subsection{The bolometric flux in scalar-tensor gravity}
\label{sec:boloflux}

We now combine the results from the previous sections and
construct the pulse profile equation in the Just spacetime,
following closely~\cite{Poutanen:2003yd,Poutanen:2006hw,Lo:2013ava,Lo:2018hes}.

Before jumping into the calculation of the bolometric flux, let us
define some more geometrical quantities that will be useful in the calculation.
Consider the radial unit vector ${\bm n}$ at the location of the
hot spot and the unit vector ${\bm k}$ along the line of sight (cf. Fig~\ref{fig:diagram}).
Since the angle $\psi$ is defined between ${\bm n}$ and the line of sight, we have
\begin{equation}
\cos\psi = {\bm k} \cdot {\bm n}\,.
\end{equation}
Due to the rotation, $({\bm k} \cdot {\bm n})$ varies periodically
\begin{equation}
\cos\psi = \cos\iota_{\rm o} \cos\theta_{\rm s}
+ \sin\iota_{\rm o}\sin\theta_{\rm s} \cos\phi_{\rm s}\,,
\label{eq:cospsi}
\end{equation}
where $\iota_{\rm o}$ is the angle between the line of
sight with respect to the spin axis, $\theta_{\rm s}$
is the spot's colatitude and $\phi_{\rm s} = 2 \pi \nu t$
is the rotational phase of the spot. We choose $t=0$ to correspond
to when the spot is closest to the observer.

The angle $\psi$ measures the {\it apparent} inclination
of the spot relative to the line of sight, but due to light-bending
this is different from the true inclination. To define the latter, recall that
${\bm k}_0$ is the unit vector in the direction in which the photon is emitted,
such that the emission angle $\alpha$ is then
\begin{equation}
\cos\alpha = {\bm k}_0 \cdot {\bm n}\,.
\end{equation}
Due to special relativistic aberration, an observer comoving with
the hot spot measures an angle $\alpha'$. The angles $\alpha$ and $\alpha'$
are related by~\cite{Poutanen:2003yd}
\begin{equation}
\cos\alpha' = \delta\, \cos\alpha\,,
\label{eq:trans_alpha}
\end{equation}
where $\delta$ is the Doppler factor
\begin{equation}
\delta = [\gamma(\theta_{\rm s}) (1 - \beta \cos\xi)]^{-1}\,.
\label{eq:doppler}
\end{equation}
Here, $\gamma(\theta_{\rm s}) = (1 - \beta^2)^{-1/2}$ is the Lorentz factor and
$\beta = v / c$ is the spot's velocity, which satisfies
\begin{equation}
\beta = \frac{2 \pi \rho_{\rm s}}{c}
\frac{\nu}{(1 - {\bar a}_{\rm s})^{b/(2a)}}
\sin\theta_{\rm s}\,.
\end{equation}
where we corrected the rotation frequency by the redshift factor
\begin{equation}
\nu / \nu_0 = \sqrt{g_{tt}(\rho_{\rm s}) / g_{tt}(\infty)}
= A_{\rm s} (1 - {\bar a}_{\rm s})^{b/(2a)}\,.
\end{equation}
The angle between ${\bm k}_0$ and the spot velocity vector ${\bm \beta}$
is denoted by $\xi$ (see Fig.~\ref{fig:diagram}), and it is related to the other angles
via~\cite{Poutanen:2006hw}:
\begin{equation}
\cos \xi =
({\bm \beta} \cdot {\bm k}_0) / \beta
= - (\sin\alpha / \sin\psi_{\rm s}) \sin\iota_{\rm o} \sin\phi_{\rm s}
\end{equation}

With this at hand, let us now return to the observed flux from the spot at energy $E$.
This quantity can be defined as
\begin{equation}
\dd F_{E} = I(E,\alpha) \dd \Omega\,,
\label{eq:flux_general}
\end{equation}
where $I(E,\alpha)$ is the specific intensity of radiation at
infinity and $\dd \Omega$ is the solid angle on the observer's sky
occupied by the spot with differential area $\dd S'$ measured
in a comoving reference frame.
Let us determine $I(E,\alpha)$ and $\dd \Omega$
separately, starting with the latter.

The solid angle can be written in terms of the impact parameter
$\sigma$ as
\begin{equation}
\dd \Omega = (\sigma\, \dd \sigma\, \dd \varphi)/D^2\,,
\end{equation}
where $D$ is the distance to the source and $\varphi$ is the azimuthal
angle around the vector ${\bm k}$ (not to be confused with the
scalar field). In terms of the surface area [cf. Eq.~\eqref{eq:just-metric}]
\begin{equation}
\dd A =
A_{\rm s}^2\rho_{\rm s}^2 (1 - {\bar a}_{\rm s})^{1-b/a} \sin\psi \dd\psi \dd\varphi\,,
\end{equation}
we have
\begin{equation}
\dd \Omega = \frac{\sigma}{A_{\rm s}^2\rho_{\rm s}^2}
\frac{(1 - {\bar a}_{\rm s})^{b/a - 1}}{\sin\psi}
\frac{\dd \sigma}{\dd \psi} \frac{\dd A}{D^2}\,.
\end{equation}
Using that the spot area projected onto the plane perpendicular to the
photon propagation direction is a Lorentz invariant, we have
\begin{equation}
\dd A' \cos\alpha' = \dd A \cos\alpha\,,
\end{equation}
where $\dd A'$ ($\dd A$) is the differential spot area measured by a comoving
(static) reference frame.
We can now write the solid angle as
\begin{equation}
\dd \Omega = \frac{\sigma}{A_{\rm s}^2 \rho_{\rm s}^2}\frac{(1 - {\bar a}_{\rm s})^{b/a -1}}{\sin\psi} \frac{\dd \sigma}{\dd \psi}
\frac{\cos\alpha'}{\cos\alpha} \frac{\dd A'}{D^2}\,,
\label{eq:solid_angle}
\end{equation}
where $dA'$ [written $(\theta,\phi)$ coordinates] is
\begin{equation}
\dd A' =
\gamma(\theta) A_{\rm s}^2\rho_{\rm s}^2 (1 - {\bar a}_{\rm s})^{1-b/a} \sin\theta \dd\theta \dd\phi\,,
\label{eq:area_comoving}
\end{equation}
where we took into account a factor of $\gamma(\theta)$ relating $\dd A$
and $\dd A'$. This term is due to Lorentz contraction of the linear
differential interval correspondent to an angular interval $\dd \phi$ measured
in the static reference frame~\cite{Lo:2018hes,Nattila:2017hdb}.

Finally, using Eq.~\eqref{eq:angle_alpha} for the impact parameter
and remembering that $\alpha$ depends implicitly on $\psi$
[cf. Eq.~\eqref{eq:psi_final}] we find:
\begin{equation}
\dd \sigma / \dd \psi = \rho_{\rm s} (1 - {\bar a}_{\rm s})^{1/2 - b/a}
(\dd \sin\alpha / \dd \psi)\,,
\end{equation}
which substituted back into Eq.~\eqref{eq:solid_angle} gives our final
expression for $\dd \Omega$:
\begin{equation}
\dd \Omega = (1 - {\bar a}_{\rm s})^{-b/a}
\cos\alpha'
\frac{\dd \cos\alpha}{\dd \cos\psi}
\frac{\dd A'}{A_{\rm s}^2 D^2} \,.
\label{eq:solid_angle_final}
\end{equation}

Now we turn our attention to the specific intensity.
We assume it can be decomposed (in the comoving reference frame)
as~\cite{Lo:2013ava}
\begin{equation}
I'_0(E'_0,\alpha') = g'(\alpha') f'(E'_0)\,,
\end{equation}
where $g'(\alpha')$ and $f'(E'_0)$ are respectively a beaming and
a spectral function, and $E'_0$ is the photon energy measured by
the comoving observer at the stellar surface.
In general, the specific intensity has a dependence on $\alpha'$ due
to Thompson scattering as the photon propagates through
the star's atmosphere~\cite{ChandraRTBook,Lo:2013ava}.

In order to express the specific intensity in terms of quantities
measured by an observer far away from the star, we proceed in two
steps. First, we Lorentz transform the energy $E'_0$ measured by
a comoving reference frame at the stellar surface to the energy
$E_0$ measured by a static reference frame near the stellar surface via
\begin{equation}
E_0 = \delta \, E'_0\,.
\label{eq:trans_mov_static}
\end{equation}
The ratio ($\textrm{specific intensity} / \textrm{energy}^3$)
is a Lorentz invariant allowing us to write
\begin{equation}
I_0(E_0, \alpha) / E^3_0
= I'_0(E'_0, \alpha') / {E'}^{3}_0
\,.
\end{equation}
We now transform $E_0$ to the energy $E$ measured by the
distant observer. These energies are related by a gravitational redshift
factor
\begin{equation}
E = A_{\rm s} (1 - {\bar a}_{\rm s})^{b/(2a)} E_0\,.
\label{eq:trans_redshift}
\end{equation}
Consequently, our final result becomes:
\begin{equation}
I(E,\alpha) = A_{\rm s}^3 \delta^3
(1 - {\bar a}_{\rm s})^{3b/(2a)}
I'_0(E'_0, \alpha')\,,
\label{eq:intens_transform}
\end{equation}
which relates the specific intensity as measured by a comoving
observer at the stellar surface to that measured by a distant observer,
including both special and general relativistic effects.

Substituting Eqs.~\eqref{eq:solid_angle_final} and~\eqref{eq:intens_transform}
into Eq.~\eqref{eq:flux_general} we obtain
\begin{equation}
\dd F_E = A_{\rm s}(1 - {\bar a}_{\rm s} )^{b/(2a)} \delta^3
I'_0(E'_0,\alpha') \cos\alpha'
\frac{\dd \cos\alpha}{\dd \cos \psi}
\frac{\dd A'}{D^2}\,.
\label{eq:spectral_flux}
\end{equation}
To obtain the bolometric flux, we integrate Eq.~\eqref{eq:spectral_flux}
over energies $E$. Using one last time the relation between $E$ and
$E'_0$ [i.e. using Eqs.~\eqref{eq:trans_mov_static}
and~\eqref{eq:trans_redshift}] we find
\begin{align}
\dd F
&\equiv \int \dd E\, \dd F_{E}\,,
\nonumber \\
&=
A_{\rm s}^2
(1 - {\bar a}_{\rm s})^{b/a}\, \delta^4\, \cos\alpha' \frac{\dd \cos\alpha}{\dd \psi} \frac{\dd A'}{D^2}
\int \dd E'_0 \,I'_0(E'_0,\alpha')\,,
\nonumber \\
&=
A_{\rm s}^2
(1 - {\bar a}_{\rm s})^{b/a}\, \delta^4\, \cos\alpha'
\frac{\dd \cos\alpha}{\dd \cos\psi} \frac{\dd A'}{D^2}I'_0(\alpha')\,.
\end{align}

As a final step, we can reexpress the angle $\alpha'$ (measured by a
comoving reference frame at the stellar surface) in terms of $\alpha$
(measured by a reference frame static near the star) using
Eq.~\eqref{eq:trans_alpha}.
Our final expression for the bolometric flux measured by an observer
due to radiation emission by a small element $\dd S'$ of the hot spot
is
\begin{equation}
\dd F =
A_{\rm s}^2 (1 - {\bar a}_{\rm s})^{b/a}\,
\delta^5\, \cos\alpha \frac{\dd \cos\alpha}{\dd \cos\psi}
\frac{\dd A'}{D^2}I'_0(\alpha')\,.
\label{eq:main_result}
\end{equation}
This formula is the main result of this paper.
It is sufficiently general to describe the Jordan-frame bolometric flux of a
static, spherically symmetric star whose exterior spacetime is described by the
Just metric, including special relativistic effects (aberration and Doppler boosts)
and gravitational redshift.

For simplicity, let us assume that the specific intensity is isotropic
(independent of $\alpha'$) and that the hot spot is small in size.
Writing $I_0(\alpha') = I'_0$, normalizing
the flux $\dd F$ by $I'_0 \dd A' / D^2$
we have\footnote{
For hot spots of large angular radius, the flux must be calculated by a discretization of
the area occupied by the spot on a grid ($\theta_i, \phi_i$), with each cell
having a corresponding Lorentz factor $\gamma(\theta_i)$, see
Eq.~\eqref{eq:area_comoving}~\cite{Lo:2013ava,Lo:2018hes}}.
\begin{equation}
F=
A_{\rm s}^2 (1 - {\bar a}_{\rm s})^{b/a}\,
\delta^5\, \cos\alpha\,
\frac{\dd \cos\alpha}{\dd \cos\psi}\,,
\label{eq:flux_normalized}
\end{equation}
which is our final result, used in Sec.~\ref{sec:results}.

When two antipodal hot spots are present, the total flux
is obtained as (see Fig.~\ref{fig:diagram})
\begin{equation}
F = F(\iota_{\rm o}, \pi - \theta_{\rm s}, \pi + \phi_{\rm s}, {\bar a}_{\rm s},
{\bar b}_{\rm s})
+ F(\iota_{\rm o}, \theta_{\rm s}, \phi_{\rm s}, {\bar a}_{\rm s},
{\bar b}_{\rm s})\,.
\label{eq:flux_normalized_antipo}
\end{equation}
The hot spot is visible for the distant observer when
$\cos\psi > \cos\psi_{\rm c}$. For the antipodal hot spot the visibility
condition is when $\cos\psi > - \cos\psi_{\rm c}$~\cite{Poutanen:2006hw}.

Let us comment on some of the features of Eq.~\eqref{eq:flux_normalized}.
First, the general form of the expression is analogous to the one found
in~\cite{Poutanen:2006hw} and agrees with it in the general relativistic limit
$a/b = 1$. The prefactor $1 - {\bar a}_{\rm s}$
reduces the peak-to-peak amplitude of the pulse's profile in comparison
to the Newtonian result (in which ${\bar a}_{\rm s} \to 0$).
Second, the Doppler factor $\delta^5$ oscillates with time, skewing the
otherwise sinusoidal pulse profile.

\section{Numerical Modeling and Pulse Profiles in Scalar-Tensor Theory}
\label{sec:results}

Having determined the expression for the radiation flux, let us
outline how the pulse profiles are numerically calculated. The steps are:
\begin{itemize}
\item [(i)] From Eq.~\eqref{eq:cospsi} we determine $\psi$ at a
given phase $\phi_{\rm s}$.
\item [(ii)] With $\psi$ at hand, we check whether or not the spot is visible. If
$\cos\psi > \cos\psi_{\rm c}$, the spot is visible and we proceed with the remaining steps.
Otherwise, the flux is zero.
\item [(iii)] With $\psi$ at hand, we then determine $\alpha$, inverting
Eq.~\eqref{eq:psi_final}. This is done numerically with the shooting method.
\item [(iv)] Knowing $\alpha$, we can calculate the derivative
$(\dd \cos \alpha / \dd \cos\psi)$ for the current value of $\psi$.
In practice, it is easier to determine
$(\dd \cos \psi / \dd \cos\alpha)^{-1}$ and evaluate it at the value of $\alpha$ found
in step (iii). To do so, we first calculate $\cos\psi$
on a fine grid $\cos\alpha \in [0, 1]$, interpolate the data using a spline
interpolation, and from this we calculate the derivative numerically.
\item [(v)] We calculate the Lorentz and Doppler factors in Eq.~\eqref{eq:doppler},
and the time-delay in Eq.~\eqref{eq:time_delay_final}.
\item [(vi)] We combine all these ingredients in Eq.~\eqref{eq:flux_normalized} to obtain
the flux. Finally, we correct the phase due to time-delay by adding $(2\pi\nu\Delta t)$
to $\phi_{\rm s}$.
\end{itemize}

Let us now show some numerical results for one hot spot with our catalog of stellar
models (see Table~\ref{tab:stellar_models}). As described in Sec.~\ref{sec:intro},
these profiles could represent observations from burst oscillations of an
accreting neutron star.
We consider two illustrative situations, one in
which $\theta_{\rm s} = \iota_{\rm o} = 45^{\circ}$ (as in~\cite{Poutanen:2006hw}),
and another with $\theta_{\rm s} = \iota_{\rm o} = 90^{\circ}$
(as in~\cite{Lo:2013ava}) to illustrate the general effects of a nonzero scalar
charge on the pulse profile. In the former case, our results agree with those presented
in~\cite{Poutanen:2006hw} in the limit of general relativity (modulo a different normalization).
To make the effects of scalar-tensor gravity more clearly visible, we consider
extreme stellar models, with $u = 0.5$ and rotation frequency
of $\nu = 600$ Hz. With these choices the strong-gravity effects on the waveform
become more pronounced in both figures.
To avoid excessive clutter in the panels and complement the sample
pulse profile shown in Fig.~\ref{fig:pp_example} we use only the models with
$A_{\rm s} > 1$ (i.e. $b_4$ and $c_4$). The waveforms for
$b_3$ and $c_3$ are similar to that shown in Fig.~\ref{fig:pp_example}.
same compactness.

\begin{figure*}[htb]
\includegraphics[width=\textwidth]{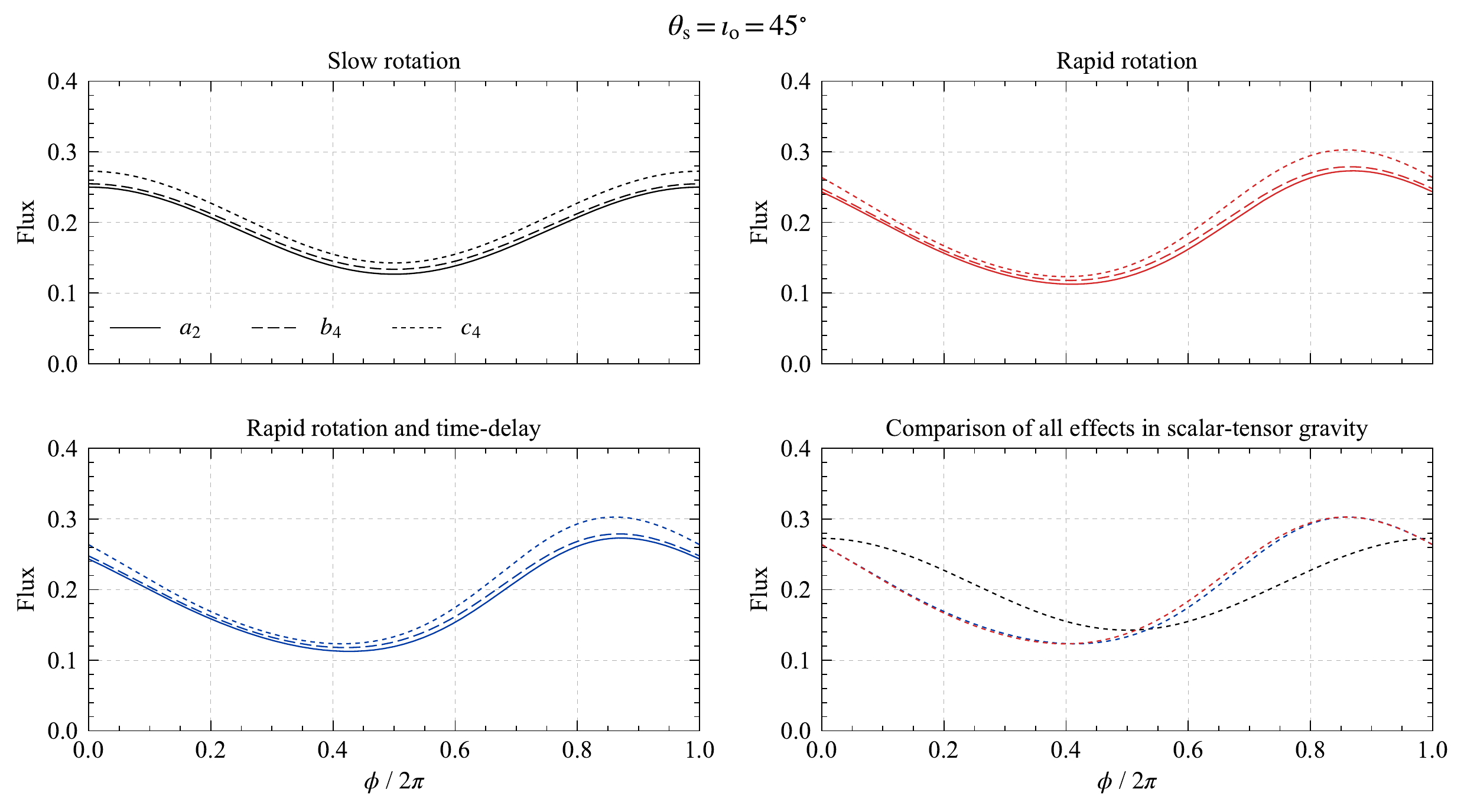}
\caption{We illustrate the effects of a nonzero scalar charge on the
pulse profile for one hot spot.
We choose $\theta_{\rm s} = \iota_{\rm o} = 45^{\circ}$ and stars
with $u = 0.5$ and increasing values of scalar charge-to-mass ration
$Q$ (models $a_2$, $b_4$ and $c_4$ in Table~\ref{tab:stellar_models}, where recall
$a_{2}$ is the general relativistic limit).
We show slowly-rotating stars ($\delta = 1$, $\Delta t = 0$) in the top-left
panel, rapidly rotating stars neglecting time-delay effects ($\delta \neq 1$, $\Delta t = 0$)
in the top-right panel, and including them ($\delta \neq 1$, $\Delta t \neq 0$)
in the bottom-left panel. In the last panel (bottom-right), we compare the
pulse profile in these three situations for a star with $Q = 1$ and $u = 0.5$
(model $c_4$ in Table~\ref{tab:stellar_models}).
}
\label{fig:illustrative_45}
\end{figure*}

Figure~\ref{fig:illustrative_45} studies the
$\theta_{\rm s} = \iota_{\rm o} = 45^{\circ}$ case, showing the pulse profile
in a number of special cases. The top-left panel considers slowly-rotating stars,
whose pulse profile is calculated setting the Doppler and Lorentz factors
to unity and neglecting the travel time delay of photons
[cf. Eq.~\eqref{eq:flux_normalized}].
The top-right (bottom-left) panel shows how the pulse profile changes when we
include Doppler effects (and time-delay). The Doppler factor changes the amplitude of
the waveform and its overall shape by skewing it. A time-delay has the effect of
shifting the arrival time of the pulse (nonuniformly over the course of a revolution),
resulting in an additional small deformation of the waveform.
Overall, the impact of each of these effects in scalar-tensor gravity is identical to
that in general relativity~\cite{Poutanen:2006hw}, the difference being
the magnitude of these effects. This is not surprising given that the flux
formula in Eq.~\eqref{eq:flux_normalized} has the same functional form in scalar-tensor
gravity as in general relativity.

\begin{figure*}[htb]
\includegraphics[width=\textwidth]{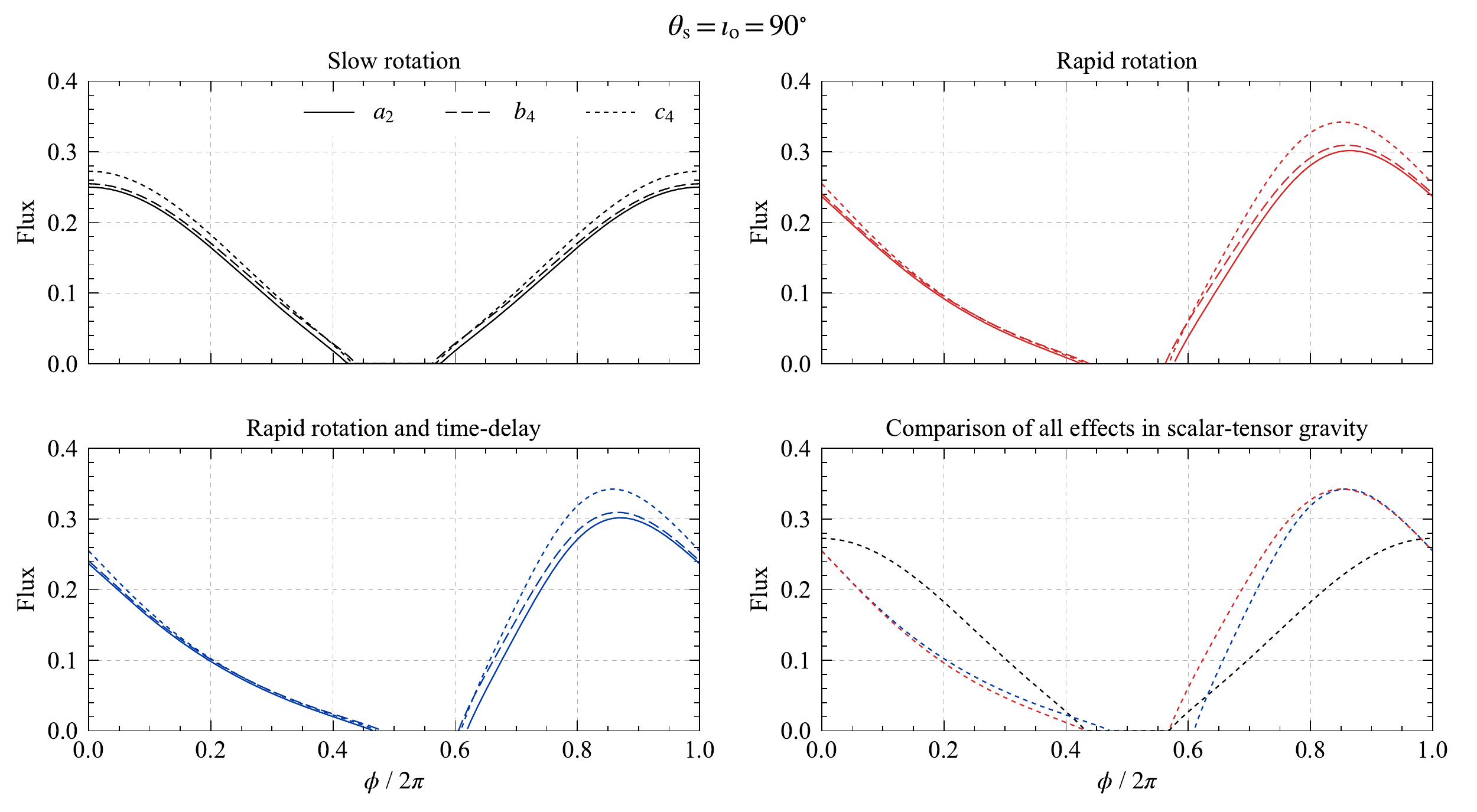}
\caption{Similar to Fig.~\ref{fig:illustrative_45}, but for
$\theta_{\rm s} = \iota_{\rm o} = 90^{\circ}$. Due to the geometrical configuration
of the hot spot and the observer, the flux hot spot becomes invisible momentarily
as the star rotates. Observe that where the flux disappears depends on the value
of $Q$.}
\label{fig:illustrative_90}
\end{figure*}

Figure~\ref{fig:illustrative_90} studies the
$\theta_{\rm s} = \iota_{\rm o} = 90^{\circ}$ case.
The same conclusions drawn from Fig.~\ref{fig:illustrative_45} are applicable here.
Because of the geometrical arrangement of the hot spot's location and the line
of sight of the observer, the hot spot becomes invisible to the observer during
certain phase intervals. These intervals when the hot spot is not visible depend
on the values of $Q$ in scalar-tensor theory, and thus, they could provide yet another
telltale sign of a deviation from general relativity. Of course, other stellar
parameters, like the inclination angle, also affect the occultation period; thus,
a full covariance analysis is necessary to determine the detectability of this effect.

\section{Conclusions and outlook}
\label{sec:conclusion}

We introduced a Just-Doppler approximation for calculating
pulse profiles of rotating neutron stars in scalar-tensor
gravity. The main result, encapsulated in Eq.~\eqref{eq:main_result},
allows one to calculate the waveforms, including effects of the strong-gravity
and special relativistic effects, generalizing the Schwarzschild plus Doppler
approximation to scalar-tensor theories of gravity. We presented a selection of
sample results for burst oscillation waveforms of an infinitesimal hot spot and
discussed the implications of the presence of a nonzero scalar charge on it.

The model independent character of the formalism opens the possibility
of constraining a large class of scalar-tensor gravity models with upcoming x-ray timing
data releases from NICER. In this regard, our work is close in spirit
to~\cite{Horbatsch:2011nh}, except that that work focused on binary pulsars. In
a forthcoming paper, we will present a statistical likelihood analysis
discussing the strength of potential future constraints on scalar-tensor gravity with
pulse profile observations. We emphasize, however, that whether future constraints
can be placed on scalar-tensor gravity with NICER data will require a full data analysis
study that varies over all model parameters in the presence of noise; such a study
is now possible thanks to the pulse-profile model presented here, and it will be carried
out in the future.

Despite the generality of the present formalism, it is necessary
to discuss its limitations and signal in which directions it
can be improved further.
At considerably large rotational frequencies ($\nu \gtrsim 300$ Hz),
the Schwarzschild-Doppler approximation becomes inappropriate
as discussed, e.g.~in~\cite{Cadeau:2006dc}, because of the rotation-induced
quadrupolar deformation of star. To include this effect in the pulse profiles, one
must first extend the Just metric to rotating stars.
To leading-order in rotation, that is, including only frame-dragging effects
while keeping a spherical geometry for the star, this calculation was done
in~\cite{Damour:1996ke}.
Using the Hartle-Thorne perturbative expansion,
Berti and Pani~\cite{Pani:2014jra} extended this work to second-order in rotation,
which includes the quadrupolar deformation of the star.
Alternatively, one could also work entirely numerically and carry out ray-tracing calculations (as
in~\cite{Cadeau:2004gm,Cadeau:2006dc,Nattila:2017hdb,Psaltis:2013zja}),
evolving the photon geodesics in a numerically constructed spacetime for rotating
neutron stars in scalar-tensor gravity~\cite{Doneva:2013qva}.
A caveat of this numerical approach is that the spacetime can only be
determined numerically {\it after} choosing a particular conformal factor $A(\vp)$
\emph{and} the equation of state, therefore inevitably making it
model-dependent.

Another approach would be to follow the work
in~\cite{Pappas:2014gca,Pappas:2015npa,Pappas:2016sye} and
use the Ernst formalism to obtain an axisymmetric spacetime
in Weyl-Papapetrou coordinates
in terms of an multipolar expansions of the multipole moments of the rotating
neutron star and the scalar field. The multipole moments appear as free constants
in the spacetime metric, thus allowing one to develop an extension
of the model-independent pulse profile model introduced here.
In this approach, and as an intermediate step, one would first have to {\it estimate} numerically
the values of these moments in order to have an analytic spacetime that
describes accurately the one constructed numerically, for example in~\cite{Doneva:2013qva}.

We emphasize, however, that Cadeau et al.~\cite{Cadeau:2006dc}
have shown that even in the presence of stellar oblateness,
the Schwarzschild-Doppler approximation works quite well at these frequencies when
$\theta_{\rm s}$ and $\iota_{\rm o}$ are {\it near the equator} as in the case
of Fig.~\ref{fig:illustrative_90}.
Geometrically, this is due to the small difference between the normal vector ${\bm n}$
on a spherical and a oblate surface near the equator, which suppresses the
effect of the star's oblateness. Because of the purely geometrical nature of
this argument, we expect the Just-Doppler approximation to accurately describe
the pulse profile when ${\theta}_{\rm s} \approx  \iota_{\rm o} \approx 90^{\circ}$
even for rapidly rotating neutron stars in scalar-tensor gravity.

Finally, it could also be interesting to consider other theories of gravity or
to use a parametrized framework -- to consider model-independent deformations of
the Tolman-Oppenheimer-Volkoff equations and the Schwarzschild spacetime -- as
introduced in~\cite{Glampedakis:2015sua,Glampedakis:2016pes}.
Work in all of these directions is currently underway and will be reported in
several future publications.

\acknowledgments
We thank George Pappas, Cole Miller, Sharon Morsink and Kent Yagi
for useful discussions and important comments on this work.
HOS thanks Lu\'is~C.~B.~Crispino, Caio~F.~B.~Macedo,
Carolina L.~Benone, Leandro~A.~Oliveira and the Universidade Federal do Par\'a for the
hospitality while part of this work was undertaken. This work was
supported by NSF Grant No. PHY-1607130 and NASA grants NNX16AB98G and
80NSSC17M0041.

\appendix

\section{Existence and location of a light ring in the Just spacetime}
\label{app:lr}

In this appendix we obtain the location for the light ring in the Just spacetime,
following closely the presentation of~\cite{Pechenick:1983}.
Consider the $(\dd \psi / \dd \rho)$ equation squared:
\begin{equation}
(\dd \psi / \dd \rho)^2
= \rho^{-4} f^{2b/a - 2} (\sigma^{-2} - \rho^{-2} f^{2b/a - 1})^{-1}.
\end{equation}
Writing $(\dd \rho / \dd \psi)^2$ instead:
\begin{equation}
(\dd \rho / \dd \psi)^2
= \rho^{4} f^{-2b/a + 2} (\sigma^{-2} - \rho^{-2} f^{2b/a - 1}).
\end{equation}
which can be rearranged as:
\begin{equation}
[\rho^{-2}(\dd \rho / \dd \psi)]^2 f^{2b/a - 2}
= \sigma^{-2} - \rho^{-2} f^{2b/a - 1}.
\end{equation}

The left-hand side is positive, so we must have
\begin{equation}
\sigma^{-2} - \rho^{-2} f^{2b/a - 1} \geq 0
\end{equation}
The second term has an extrema at some $\rho_{\rm c}$ for which:
\begin{equation}
\dd (\rho^{-2} f^{2b/a - 1}) / \dd \rho = 0.
\end{equation}
Taking the derivative, we find that it occurs at
\begin{equation}
\rho_{\rm c} = \frac{1}{2}(a + 2b).
\end{equation}
In the general relativistic limit, we have that $a = 2 G m / c^2 = b$, and we
recover the usual location of the light ring in the Schwarzschild spacetime, i.e.
$3 G M / c^2$.

In the Just metric, for the light ring to be
within the star, the star's radius $\rho_{\rm s}$ must satisfy:
\begin{equation}
\rho_{\rm s} \geq \frac{1}{2}(a + 2b)
%
= \frac{G_{\ast} m}{c^2} ( 2 + \sqrt{1 + Q^2} )
\end{equation}
Equivalent expressions can be obtained dividing the inequality by $\rho_{\rm s}$ and
with a few rearrangements:
\begin{equation}
{\bar a}_{\rm s} \leq 2(1 - {\bar b}_{\rm s})\,, \quad
({\bar b}_{\rm s}/2) (2 + \sqrt{1+Q^2} ) \leq 1\,.
\end{equation}
These equalities are shown in Fig.~\ref{fig:existence} by the dot-dashed lines.

\bibliographystyle{apsrev4-1}
\bibliography{biblio}

\end{document}